\documentclass[aps,a4paper, preprint, superscriptaddress,preprintnumbers,floatfix,nofootinbib,amsmath,amssymb]{revtex4-1}
\usepackage{multirow}
\usepackage{url}
\usepackage{hyperref}
\usepackage{color}
\usepackage{cancel}
\usepackage{cleveref}
\usepackage{subfig}
\usepackage{soul}
\usepackage[normalem]{ulem}

\usepackage{amstext,amssymb}
\usepackage{amsmath}
\usepackage{graphicx}
\usepackage{diagbox}
\usepackage{slashbox}
\usepackage{slashbox}
\usepackage{xspace}
\usepackage{color}
\usepackage{units}
\usepackage[T1]{fontenc}
\usepackage{amsmath,bm}
\usepackage{nicefrac}
\usepackage{slashed} 
\usepackage{multirow}
\newcommand{\be}{\begin{equation}}
\newcommand{\ee}{\end{equation}}
\newcommand{\bea}{\begin{eqnarray}}
\newcommand{\eea}{\end{eqnarray}}

\usepackage{slashed}

\newcommand{\nua}[1]{\ensuremath{\rlap{\kern-2.5pt\ensuremath{\overset{\scriptscriptstyle(-)}{\phantom{\nu}}}}{\ensuremath{{\nu}_{#1}}}}\xspace}

\newcommand{\deltaCP}{\ensuremath{\delta_{\rm CP}}}

\definecolor{brickred}{rgb}{0.8, 0.25, 0.33}
\definecolor{brightcerulean}{rgb}{0.11, 0.67, 0.84}
\definecolor{brown(traditional)}{rgb}{0.59, 0.29, 0.0}
\begin{document}
\title{Probing non-unitarity of the PMNS matrix in P2SO and comparison with DUNE}

\author{Sambit Kumar Pusty}
\email{pustysambit@gmail.com}
\affiliation{School of Physics,  University of Hyderabad, Hyderabad - 500046,  India}

\author{Samiran Roy}
\email{samiranroy.hri@gmail.com}
\affiliation{School of Physics,  University of Hyderabad, Hyderabad - 500046,  India}

\author{Monojit Ghosh}
\email{mghosh@irb.hr}
\affiliation{Center of Excellence for Advanced Materials and Sensing Devices, Ru{\dj}er Bo\v{s}kovi\'c Institute, 10000 Zagreb, Croatia}             
\author{Rukmani Mohanta}
\email{rmsp@uohyd.ac.in}
\affiliation{School of Physics,  University of Hyderabad, Hyderabad - 500046,  India}

\begin{abstract}

We compare the sensitivity of the upcoming long-baseline neutrino experiments Protvino to Super-ORCA (P2SO) and the Deep Underground Neutrino Experiment (DUNE) to non-unitarity (NU) of the leptonic mixing matrix in a model-independent framework. NU can arise in theories beyond the Standard Model that include heavy neutral leptons. These effects can modify neutrino oscillation probabilities and introduce new sources of CP violation, which may affect precision measurements of neutrino parameters. We find that DUNE provides stronger bounds on $\alpha_{11}$ and $|\alpha_{21}|$, while P2SO shows better sensitivity to $\alpha_{22}$ and $\alpha_{33}$, mainly due to its longer baseline and stronger matter effects. Our results show that DUNE (P2SO) will be able to improve the current bounds of $\alpha_{11}$ ($\alpha_{33}$). We further examine correlations with standard oscillation parameters and quantify the impact of NU on mass hierarchy, octant, and CP-violation sensitivities. Our results show that these sensitivities depend upon NU in a non-trivial way interconnecting the parameter degeneracies and matter effects. Our results demonstrate the complementarity of P2SO and DUNE in probing NU and show that NU can significantly influence next-generation precision oscillation studies.
 
\end{abstract}

\maketitle
\flushbottom


\section{INTRODUCTION}

The discovery of neutrino oscillation provides the first conclusive evidence of physics beyond the Standard Model (BSM) and confirmed that neutrinos have a small but nonzero mass. This groundbreaking finding challenged the initial assumption of massless neutrinos in the Standard Model~\cite{Kajita:2016cak,McDonald:2016ixn}. The conventional three-flavor oscillation scenario is delineated by three mixing angles ($\theta_{12}$, $\theta_{13}$, and $\theta_{23}$), two mass-squared  differences ($\Delta m^2_{21}$ and $\Delta m^2_{31}$), and a CP violation phase ($\deltaCP$). Out of the three mixing angles, the value of the atmospheric angle $\theta_{23}$  still remains obscure and ambiguous  whether it lies in the higher octant ($\theta_{23}>45^\circ$) or lower octant ($\theta_{23}<45^\circ$). The results from  current experiments hint towards higher octant~\cite{NOvA:2021nfi,T2K:2025yoy,IceCubeCollaboration:2024ssx}; however, greater accuracy is needed. Another ambiguity persists in the atmospheric mass splitting term $\Delta m^2_{31} = m_3^2 - m_1^2$. It remains unclear if the value is greater (normal hierarchy)  or less (inverted hierarchy) than zero. Regarding the CP-violating phase $\deltaCP$, it is still one of the most unsettled issues in the neutrino field. It is crucial to ascertain the CP violating phase since it offers an indication to answer the matter-antimatter asymmetry of the universe through leptogenesis~\cite{Fukugita:1986hr,Joshipura:2001ui,Endoh:2002wm,Buchmuller:2005eh}.
The most recent results from the joint fit of T2K and NO$\nu$A point towards the value of the CP-violating phase $\deltaCP \approx-90^{\circ}$ in the inverted ordering, but they can not completely rule out the possibility of CP-conserving values in the normal ordering~\cite{T2K:2025wet}. 

Neutrino oscillation experiments have precisely measured the mass-squared differences, providing strong evidence that neutrinos possess mass. However, the absolute mass scale of neutrinos remains unknown, as oscillation experiments are sensitive only to mass differences and not the exact masses of individual neutrino states. Determining this absolute mass scale remains a key open question in neutrino physics. Direct searches, such as the KATRIN experiment, provide upper limits on neutrino mass ($m_{\nu_e}<0.45~$eV)~\cite{KATRIN:2024cdt}, while indirect constraints from cosmology suggest that the sum of neutrino masses is less than 0.12 eV \cite{DESI:2024hhd,Shao:2024mag,DESI:2025gwf}. Despite these efforts, the mechanism by which neutrinos acquire mass is still a mystery. Unlike other Standard Model fermions, neutrinos could be either Dirac particles, acquiring mass via a Yukawa coupling, or Majorana particles, whose mass could arise from the seesaw mechanism, which introduces new heavy neutrino states.

Many extensions of the SM appeared in the literature that describe the mechanism of mass generation in the neutrino sector through the inclusion of new heavy particles, commonly referred to as seesaw mechanisms.  For example,  type-I seesaw~\cite{Mohapatra:1979ia,Gell-Mann:1979vob,Schechter:1980gr,Valle:2015pba} that uses heavy neutral leptons (HNLs)  generally in the  scale ${\cal O}(10^{14})$ GeV to account for the lightness of the active  neutrino masses in the sub-eV scale. Due to the massive nature of the HNLs, they do not directly participate in the neutrino oscillation phenomena, rather mix with the active three-flavour neutrinos, leading to the non-unitarity (NU) in the $3 \times 3$ leptonic mixing matrix or the so-called Pontecorvo–Maki–Nakagawa–Sakata (PMNS) matrix \cite{Maki:1962mu}. Even with LHC experiments, it is challenging to probe the high-scale seesaws. On the other hand, low-scale seesaws are intriguing due to their possible experimental signature.   The new neutral lepton states with masses of the order of GeV/TeV scale, commonly known as sterile neutrinos, are present in low-scale seesaw models, such as inverse seesaw, linear seesaw, etc., which can bring down the seesaw  scale to TeV range  and lack any SM interaction. It should be further noted that, in addition to the unknowns in the three flavour neutrino mixing, 
there are several anomalies from short-baseline experiments such as MiniBooNE~\cite{MiniBooNE:2018esg} and LSND~\cite{LSND:2001aii}, alongside reactor anomalies~\cite{Mention:2011rk,Huber:2011wv}. These anomalies are best explained by the existence of new sterile neutrino states, which can be on the order of eV or less \cite{Abazajian:2012ys}. If there are such additional iso-singlet new states in the standard model extensions, they will alter the $3\times3$ neutrino mixing matrix via influencing the three flavor oscillation and in the averaged-out regime, the oscillation probability effectively leads to NU at leading order~\cite{Blennow:2016jkn}.
The majority of HNLs possess masses significantly larger than those of active neutrinos, preventing them from being produced through the same mechanisms as active neutrinos and kinematic constraints prevent HNLs from transitioning into active states. The theoretical model and its implications on neutrino oscillations have been studied in great detail in the literature~\cite{Antusch:2006vwa,Escrihuela:2015wra,Blennow:2025qgd}.
Additionally, NU effects have been explored in a number of experiments~~\cite{Gronau:1984ct,
  Nardi:1994iv,Goswami:2008mi, Atre:2009rg,Escrihuela:2016ube,Fernandez-Martinez:2016lgt,Blennow:2023mqx,Forero:2021azc,Denton:2021mso,Dutta:2019hmb,Ge:2016xya,Meloni:2009cg,Miranda:2018yym,C:2017scx,Miranda:2020syh,Coloma:2021uhq,Agarwalla:2021owd} in recent years.
  
NU effects can introduce new CP-violating phases, alter oscillation probabilities, and distort measurements of standard oscillation parameters such as the neutrino mass ordering or the leptonic CP phase $\deltaCP$.  It is important to note that, the additional phases due to NU  appear in combination with $\deltaCP$ and hence,  the NU scenario will hamper the CP violation sensitivity of the oscillation experiments. For precision measurements in the upcoming   experiments to be robust, it is crucial to quantify and constrain these parameters. Long-baseline (LBL) experiments offer a great opportunity for such investigations. Neutrino oscillations are sensitive to matter effects, which are amplified when propagation occurs over long distances. In this study, we focus on two LBL experiments: Deep Underground Neutrino Experiment (DUNE) and Protvino to Super ORCA (P2SO) to study NU in the PMNS matrix. DUNE  has a wideband high-intensity beam with a baseline of 1300 kms. The liquid argon TPC detector offers excellent event reconstruction, and the strong matter effects give DUNE high sensitivity to neutrino parameters. Study of NU in the context of DUNE has been studied extensively \cite{Escrihuela:2016ube,Dutta:2016czj,Hernandez-Garcia:2017pwx,Dutta:2019hmb,Chatterjee:2021xyu,Agarwalla:2021owd,Trzeciak:2025hap}. P2SO, with its proposed 2595 km baseline, probes a distinct energy range in the few-GeV region and experiences different matter density profiles. Using complementary baselines, energy coverage, and detector technology, NU searches in these two scenarios provide a more reliable analysis of NU on precision oscillation physics.
 Previously Ref.~\cite{Kaur:2021rau} studied the NU in the context of Protvino to ORCA (P2O) experiment using the same baseline. However, our work is novel from this study is several ways. First of all, in our P2SO configuration, we have considered improved beam power and updated Super-ORCA detector. It was shown in Ref.~\cite{Singha:2021jkn}, though this experiment has the longest baseline among the current and future generation accelerator based experiments, its sensitivity is suppressed in the P2O configuration due to large backgrounds. This is due to the fact that the ORCA detector is not optimized for the neutrinos coming from the beam. Therefore, the P2O configuration is limited to explore the full potential of the 2595 km baseline which is also known as the bi-magic baseline \cite{Raut:2009jj,Dighe:2010js}. However, the P2SO configuration is well suited  to study a physics phenomena in presence of the large-matter effect. So, even though the P2SO experiment is proposed for the next decade, it is extremely important for academic purposes to understand the behaviour of the NU with stronger matter effect and pave the path for the future low energy neutrino factories \cite{Denton:2025kvy}. In addition, Ref.~\cite{Kaur:2021rau} considers only one parameter at a time, whereas our study also includes the case when all the parameters are taken at the same time. Furthermore, our work explores, non-trivial correlations between standard and NU parameters and the effect of the phase of the off-diagonal NU parameter $\alpha_{21}$. Finally, our treatment of NU in determining physics sensitivities is different as compared to Ref.~\cite{Kaur:2021rau}. We will comment on this in section~\ref{EMH}.
The goal of this work is to study the sensitivity of P2SO to NU in a model independent way~\cite{Escrihuela:2015wra} and compare its sensitivity with DUNE. This allows us to comprehend the behavior of NU in an experiment with stronger matter effect as compared to DUNE.

The paper is organized as follows: Sec. \ref{Sec: Form} outlines the theoretical formalism in the presence of NU, while Sec. \ref{analytical} presents the analytical expressions considering NU effects. Sec. \ref{exp} and \ref{sim} detail the simulations performed in this study, along with the experimental specifications of DUNE and P2SO. Finally, Sec. \ref{res} discusses the results, followed by the conclusions in Sec. \ref{conclusion}.


\section{Formalism}
\label{Sec: Form}

The mixing of the heavy sterile states with the standard active neutrinos  makes the $3 \times 3$ leptonic mixing matrix non-unitary.  Model-independent parameterization of the NU mixing matrix can be achieved due to the large masses of the heavy sterile neutrinos.
Those heavy states cannot be directly produced or take part in the neutrino oscillation because of their large masses \cite{Xing:2011ur, Fernandez-Martinez:2007iaa, Escrihuela:2015wra}. Generally, the number of new particles has no impact on this parameterization \cite{Escrihuela:2015wra}.  The full mixing matrix is  given by a unitary mixing matrix ($U^{n\times n}$), having four sub-matrices as~\cite{Hettmansperger:2011bt}

\begin{equation}
 U^{n\times n} = 
 \begin{pmatrix}
  N^{3\times 3} & S^{(n-3)\times n}
  \\
  V^{n\times (n-3)} & T^{(n-3)\times (n-3)}
 \end{pmatrix}\,.
\label{eq:Unxn}
\end{equation}
Here, $n$ is the total number of neutrino states, out of which 3 are light active states and the rest ($n-3$) are the new states coming from the extension of the SM.  $S$ and $V$ represent the mixing between the light  and the new heavy states, whereas the $T$ matrix describes the self-mixing among the new states of the neutrinos.
Clearly, from the unitarity condition, the sub-matrices can take the form
\begin{eqnarray}
 NN^\dagger + SS^\dagger = I , \nonumber \\
 TT^\dagger + VV^\dagger = I . 
\end{eqnarray}
Therefore, the $N$ matrix $(3 \times 3)$  describing the mixing among the three light neutrino states is no longer unitary. The NU   sub-matrix can be conveniently parameterized by multiplying a triangular matrix by the standard three flavor neutrino mixing matrix. It can be represented in the following manner~\cite{Escrihuela:2015wra}
\begin{equation}
 N = N^{NP}U=
 \begin{pmatrix}
  \alpha_{11} & 0 & 0
  \\
  \alpha_{21} & \alpha_{22} & 0
  \\
  \alpha_{31} & \alpha_{32} & \alpha_{33}
 \end{pmatrix}U\,.
\label{Nmatrix}
\end{equation}
Here, all of the NU effects are encoded in the lower triangular matrix $N^{NP}$. All diagonal parameters $\alpha_{ii}$ are real and close to one, while the off-diagonals $\alpha_{ij}$ ($i\neq j$) are complex and have very small magnitudes. The complex off-diagonal parameters are associated with the corresponding arguments $\phi_{ij}$ ($i\neq j$), which contribute as new sources for CP violation. In the limiting case, when $N^{NP}$ becomes identity ($N^{NP}=I$), we get back our standard PMNS matrix, $U$. 

It is worthy to mention that NU effects can also be illustrated through alternative parameterization, such as defining the light neutrino mixing matrix as 
$N=(1- \eta)U'$, a form commonly employed in the study of NU scenarios \cite{Fernandez-Martinez:2007iaa}. Both $U$ and $U'$ are unitary matrices that are equivalent to the standard PMNS matrix, differing only by small corrections proportional to the terms  $\eta$ and $\alpha$~\cite{Blennow:2016jkn}. In this approach,  $\eta$ is a Hermitian 3×3 matrix given by,
\begin{equation}
 \eta = 
 \begin{pmatrix}
  \eta_{11} & \eta_{12}& \eta_{13}
  \\
  \eta_{12}^* & \eta_{22} &\eta_{23}
  \\
  \eta_{13}^* & \eta_{23}^* & \eta_{33}
 \end{pmatrix}\,,
\label{Nmatrix}
\end{equation}
which characterizes the extent of unitarity violation in the mixing framework. 

In our work, we use the triangular representation of the NU. The connection between these two parameterizations is further described in~\cite{Blennow:2016jkn} in great detail.


\section{Analytical Expressions}
\label{analytical}

In the presence of NU, the vacuum transition probability from a flavor neutrino $\nu_{\alpha}$ to another flavor neutrino $\nu_{\beta}$ can be calculated using the following equation \cite{Escrihuela:2015wra},
\begin{align}
\label{general}
 P_{\alpha \beta}&=\sum^3_{i,j}N^{*}_{\alpha i}N_{\beta i}N_{\alpha j}N^{*}_{\beta j} -4 \sum^3_{j>i}\operatorname{Re}\left[N^{*}_{\alpha j}N_{\beta j}N_{\alpha i}N^{*}_{\beta i}\right]\sin^2 \left(\frac{\Delta m^2_{ji}L}{4E_{\nu}}\right) \\
 & ~~~~~~~~~~~~~~~~+2 \sum^3_{j>i}\operatorname{Im}\left[N^{*}_{\alpha j}N_{\beta j}N_{\alpha i}N^{*}_{\beta i}\right]\sin \left(\frac{\Delta m^2_{ji}L}{2E_{\nu}}\right). \notag
\end{align}

Here, $\Delta m^2_{ji}$ are the mass-squared differences ($\Delta m^2_{21}$ and $\Delta m^2_{31}$), $E_{\nu}$ represents energy of the propagating neutrino along the baseline of length $L$.
Now, using Eq.~\ref{general}, the transition probability for the electron neutrino appearance ($P_{\mu e}$) is given by
\begin{align}
& P_{\mu e} = \alpha_{11}^2 |\alpha_{21}|^2 -4
        \sum\limits^3_{j>i} Re\left[ 
        N^*_{\mu j}N_{ej}N_{\mu i}N^*_{ei} \right]
        \sin^2\left( \frac{\Delta m^2_{ji}L}{4E}\right) \nonumber\\
&+ 
        2 \sum\limits^3_{j>i} Im\left[
        N^*_{\mu j}N_{ej}N_{\mu i}N^*_{ei}\right] 
        \sin\left(\frac{\Delta m^2_{ji}L}{2E}\right)  \,.
\end{align}
Using appropriate approximations by ignoring the small cubic terms of ${\alpha_{21}}$, $\sin \theta_{13}$, and $\Delta m^2_{21}$, we get a simple expression for $P_{\mu e} $ as follows~\cite{Escrihuela:2015wra}, 
\begin{equation}
\label{eq:P_mue}
P_{\mu e} = \alpha_{11}^2 |\alpha_{21}|^2+\alpha_{11}^2\alpha_{22}^2 P^{3\times 3}_{\mu e}+\alpha_{11}^2\alpha_{22} |\alpha_{21}|P^{I}_{\mu e}\;,
\end{equation} 
where $P^{3\times 3}_{\mu e}$ is the standard oscillation probability and $P^{I}_{\mu e}$ is the oscillation probability that contains the effects of NU in the form of new CP phase, $\phi_{21}$. The  precise forms of $P^{3\times 3}_{\mu e}$ and $P^{I}_{\mu e}$ are shown below,
\begin{align}
\tiny
P^{3\times 3}_{\mu e} &= 
 4 \bigg[\cos^2\theta_{12} \cos^2\theta_{23} 
   \sin^2\theta_{12} \sin^2\left(\frac{\Delta m^2_{21}L}{4E}\right) 
 +  \cos^2\theta_{13}\sin^2\theta_{13}
   \sin^2\theta_{23}\sin^2\left(\frac{\Delta m^2_{31}L}{4E}\right) \bigg] \nonumber\\ 
&+ 
  \sin 2\theta_{12}
   \sin\theta_{13}\sin 2\theta_{23}
   \sin\left(\frac{\Delta m^2_{21}L}{2E}\right) 
\sin\left(\frac{\Delta m^2_{31}L}{4E}\right) 
   \cos\left(  \delta_\mathrm{CP}+\frac{\Delta m^2_{31}L}{4E} \right)  \,,  
\end{align}
\begin{align}
P^{I}_{\mu e}  &= 
-2 
   \bigg[
   \sin(2\theta_{13}) \sin\theta_{23} 
   \sin\left( \frac{\Delta m^2_{31}L} {4E}\right)
  \sin\left( \delta_\mathrm{CP} - \phi_{21}+\frac{\Delta m^2_{31}L}{4E}  \right) \bigg]
\nonumber \\ 
  & +   \cos\theta_{13} \cos\theta_{23} 
  \sin 2\theta_{12} \sin \phi_{21}
   \sin\left(\frac{\Delta m^2_{21}L}{2E}\right)
   .
\end{align}
Analogously, one can write the disappearance probability  $P_{\mu \mu}$ similar to Eq. (\ref{eq:P_mue}) as,
\begin{equation}\label{eq:P_mumu}
 P_{\mu \mu} = \alpha_{22}^{4}P_{\mu \mu}^{3\times 3} + \alpha_{22}^{3}|\alpha_{21}|P_{\mu \mu}^{I_{1}} + 2\alpha_{22}^{2}|\alpha_{21}|^{2}P_{\mu \mu}^{I_{2}}\;,      
\end{equation}
with $P_{\mu \mu}^{3\times 3}$ being the standard three-flavor probability, with an approximate expression given by,

\begin{equation}
 \begin{split}
& P_{\mu \mu}^{3\times 3}  \approx 1-4 \left[\cos^2\theta_{23} 
\sin^2\theta_{23}-\cos(2 \theta_{23}) \sin^2\theta_{23} \sin^2{\theta_{13}} 
\right] \,\sin^2 {\left(\frac{\Delta m^2_{31}L}{4E} \right)}\\
&+2 \left[\cos^2\theta_{12} \cos^2\theta_{23} 
\sin^2\theta_{23}- \cos(\text{I}_{123}) \cos \theta_{23} \sin 
(2 \theta_{12}) \sin^3 \theta_{23} \sin{\theta_{13}} 
\right]\,   \sin{\left(\frac{\Delta m^2_{31}L}{2E} \right)} \,\sin{\left(\frac{\Delta m^2_{21}L}{2E} \right)}\\
&-4 \left[\cos^2\theta_{12} \cos^2\theta_{23} 
\sin^2\theta_{23} \,\cos{\left(\frac{\Delta m^2_{31}L}{2E} \right)} +  
\cos^2\theta_{12} \cos^4\theta_{23} 
\sin^2\theta_{12} \right]\,\sin^2{\left(\frac{\Delta m^2_{21}L}{4E}
\right)} \, ,
 \end{split}
\end{equation}
%

The terms $P_{\mu \mu}^{I_{1}}$ and $P_{\mu \mu}^{I_{2}}$ are expressed as \cite{Escrihuela:2015wra},
\begin{eqnarray}
P_{\mu \mu}^{I_{1}}  & \approx &  -8\left[ \sin\theta_{13}\sin\theta_{23}\cos(2\theta_{23})\cos(\delta_{CP} +\phi_{21})\right] \sin^2 {\left(\frac{\Delta m^2_{31}L}{4E} \right)} \nonumber \\
&+& 2\left[\cos\theta_{23}\sin(2\theta_{12})\sin^{2}\theta_{23}\cos(\phi_{21})\right]\sin{\left(\frac{\Delta m^2_{31}L}{2E} \right)} \sin{\left(\frac{\Delta m^2_{21}L}{2E} \right)},
\end{eqnarray}
and
\begin{equation}
P_{\mu \mu}^{I_{2}} \approx 1 - 2\sin^{2}\theta_{23}\sin^{2}{\left(\frac{\Delta m^2_{31}L}{4E} \right)}.
\label{eq.14}
\end{equation}

In addition to the standard oscillation parameters, the above probabilities depend explicitly on $\alpha_{11}$, $\alpha_{22}$, and $\alpha_{21}$, while the remaining NU parameters influence the probabilities only through their contribution to matter effects \cite{Blennow:2016jkn, Dutta:2016czj}.


\section{Experimental Details}
\label{exp}

The two forthcoming long-baseline experiments, P2SO and DUNE, are the primary focus of this investigation. The essential experimental details are summarized as follows.

\subsection{P2SO}

 The  P2SO is a  forthcoming long-baseline experiment. The neutrino source will be located at Protvino, Russia, which is a U-70 synchrotron.  The source is then directed toward the detector located in the Mediterranean Sea, about 40 km off the coast of Toulon, France.
 The  baseline for this experiment is 2595 km.  The detailed description    of the experiment can be found in Refs.~\cite{Akindinov:2019flp, Singha:2022btw,Majhi:2022fed,Singha:2023set}. For the P2SO experiment, the energy window spans from 0.2 GeV to 10 GeV, peaking at about 5 GeV. The accelerator will produce a 450 KW beam corresponding to $4 \times 10^{20}$ protons on target annually for this configuration. Compared to the ORCA, the Super-ORCA detector will be 10 times more dense.  We have considered a total run period of six years consisting of three years in neutrino and three years in antineutrino modes. 


\subsection{DUNE}
The DUNE is an upcoming long-baseline neutrino oscillation experiment,   located at Fermilab. It consists of a far detector that is a  40 kt liquid argon time projection chamber (LArTPC) located in South Dakota.
In order to simulate the DUNE experiment, we use the official technical design report (TDR) \cite{DUNE:2021cuw} GLoBES files. 
Additionally, we consider the total run-time to be thirteen years, involving six and a half years in neutrino mode and six and a half years in antineutrino mode. This period is equivalent to an accumulation of $1.1 \times 10^{21}$ protons on target (POT) per year. The DUNE experiment features a high beam power of 1.2 MW and performs over an extensive band of neutrino energies. This configuration is equivalent to ten years of data taking, based on the standard staging assumptions described in \cite{DUNE:2020jqi}.

\section{Simulation Details}
\label{sim}
We have used the GLoBES \cite{Huber:2004ka, Huber:2007ji} software tool to simulate P2SO and DUNE experiments. To integrate NU effect, we have made modifications to the GLoBES probability engine. The sensitivity has been estimated using the Poisson log-likelihood formula:

\begin{equation}
 \chi^2_{{\rm stat}} = 2 \sum_{i=1}^n \bigg[ N^{{\rm test}}_i - N^{{\rm true}}_i - N^{{\rm true}}_i \log\bigg(\frac{N^{{\rm test}}_i}{N^{{\rm true}}_i}\bigg) \bigg]\,,
\end{equation}
where $N^{{\rm test}}$ and $N^{{\rm true}}$ represent the number of events in the test and true spectra, respectively, and $n$ denotes the number of energy bins. The systematic error is incorporated by the method of pull~\cite{Fogli:2002pt,Huber:2002mx}. Table~\ref{spara} displays the values of the oscillation parameters, which are taken from NuFit 6.0. For the true oscillation parameters, we consider the central values of these best-fit parameters. During the $\chi^2$ analysis, we consider one NU parameter at a time for simplicity unless otherwise mentioned.
\begin{table}[t] 

    \centering
    \begin{tabular}{|c|c|c|} \hline 
         Parameters&  Best-fit value$ \pm 1\sigma$ & $3\sigma$\\ \hline 
         $\sin^2\theta_{12}$&  $0.308^{+0.012}_{-0.011}$& 0.275 → 0.345\\ \hline 
         $\sin^2\theta_{13}$&  $0.02215^{+0.00056}_{-0.00058}$& 0.02030 → 0.02388\\ \hline 
         $\sin^2\theta_{23}$&  $0.470^{+0.017}_{-0.013}$& 0.435 → 0.585\\ \hline 
         $\delta_{CP} ~[^{\circ}]$&  $212^{+26}_{-41}$& 124 → 364\\ \hline 
         $\Delta m_{21}^2/ 10^{-5}~{\rm eV}^2$&  $7.49^{+0.19}_{-0.19}$& 6.92 → 8.05\\ \hline 
         $\Delta m_{31}^2/ 10^{-3}~{\rm eV}^2$&  $+2.513^{+0.021}_{-0.019}$& 2.451 → 2.578\\ \hline
    \end{tabular}
    \caption{True oscillation parameters used in our analysis,  is taken from NuFIT 6.0~\cite{Esteban:2024eli} considering  normal mass hierarchy. }
    \label{spara}
\end{table}  

\section{Results}
\label{res}

In this section, we discuss the primary results highlighting how NU parameters affect the physics sensitivities of the proposed long-baseline experiments DUNE and P2SO. Among six NU parameters, we will focus on $\alpha_{11}$, $\alpha_{22}$, $\alpha_{33}$ and one off-diagonal parameter $\alpha_{21}$, which also includes the associated CP-violating phase $\phi_{21}$. Other off-diagonal terms ($\alpha_{31}$ and $\alpha_{32}$) have negligible effects on the appearance and disappearance channel probabilities within the current limits on those parameters \cite{Dutta:2019hmb}. Hence, we exclude these parameters from our analysis. We illustrate how the presence of NU parameters $\alpha_{11}$, $\alpha_{22}$, $\alpha_{33}$, and $\alpha_{21}$ affect the standard neutrino oscillation paradigm and put limits on these  parameters based on the upcoming DUNE and P2SO experiments.  We have considered the normal mass hierarchy throughout our simulations. The oscillation parameters $\theta_{23}$ and $\Delta m_{31}^2$ are marginalized  within their respective $3\sigma$ allowed ranges. The CP phase $\deltaCP$ is free and marginalized over its full range, \textit{i.e.}, $\delta_{CP}\in [0, 360^\circ]$. All other standard oscillation parameters are fixed at their true values. The new CP phase $\phi_{21}$ is also marginalized over the full range of $ 0$ to $ 360^\circ$. In the standard neutrino oscillation scenario, each diagonal parameter is assigned a value of 1 ($\alpha_{ii} = 1$), while the off-diagonal parameters are set to zero ($\alpha_{ij} = 0$, for $i \neq j$).

\subsection{Sensitivity limits}

\begin{figure}[h]
\begin{center}
    \includegraphics[width=75mm, height=62mm]{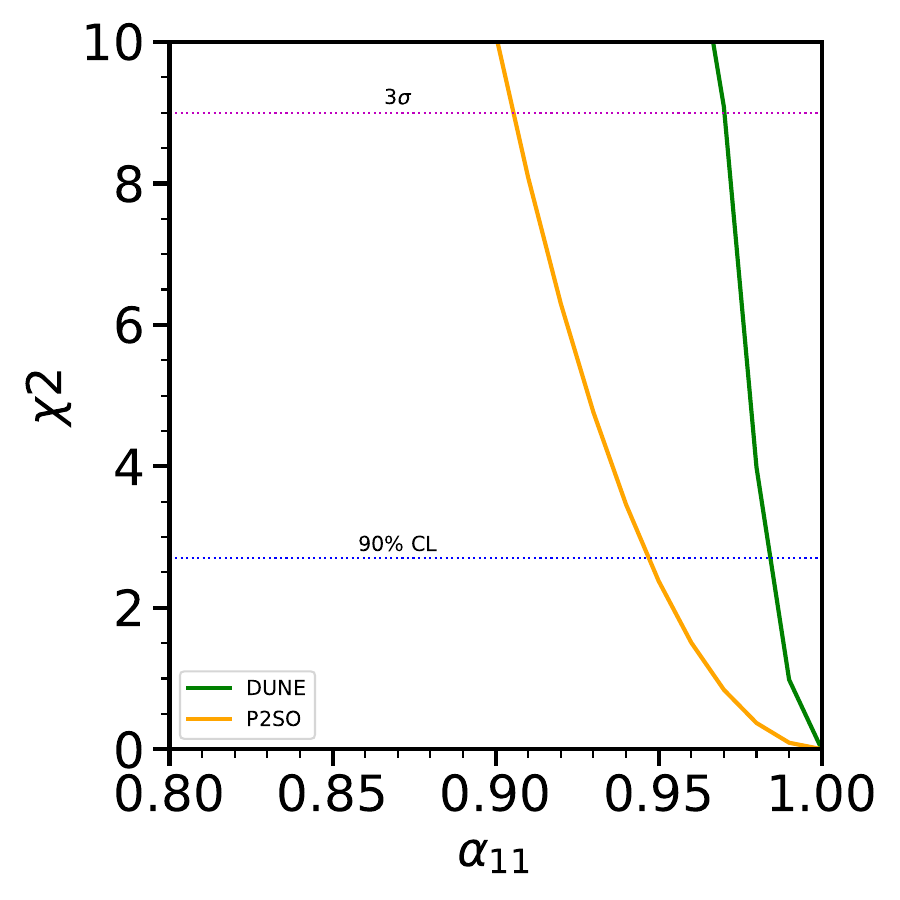}
    \includegraphics[width=75mm, height=62mm]{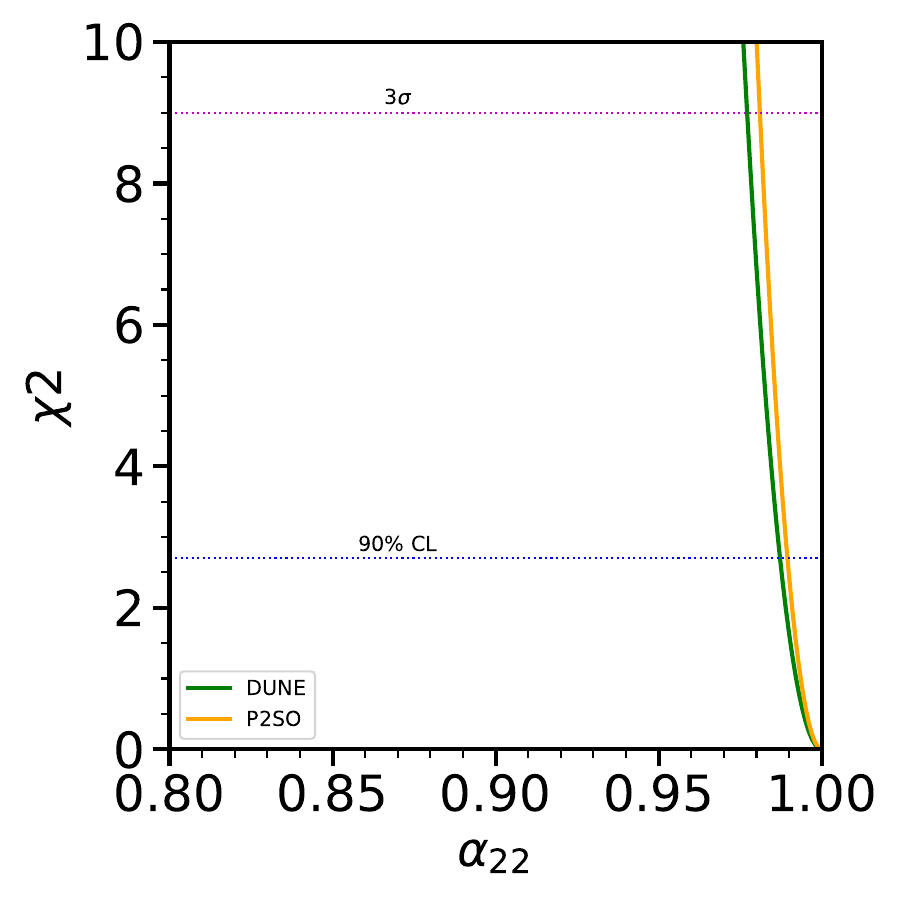}  
    \includegraphics[width=75mm, height=62mm]{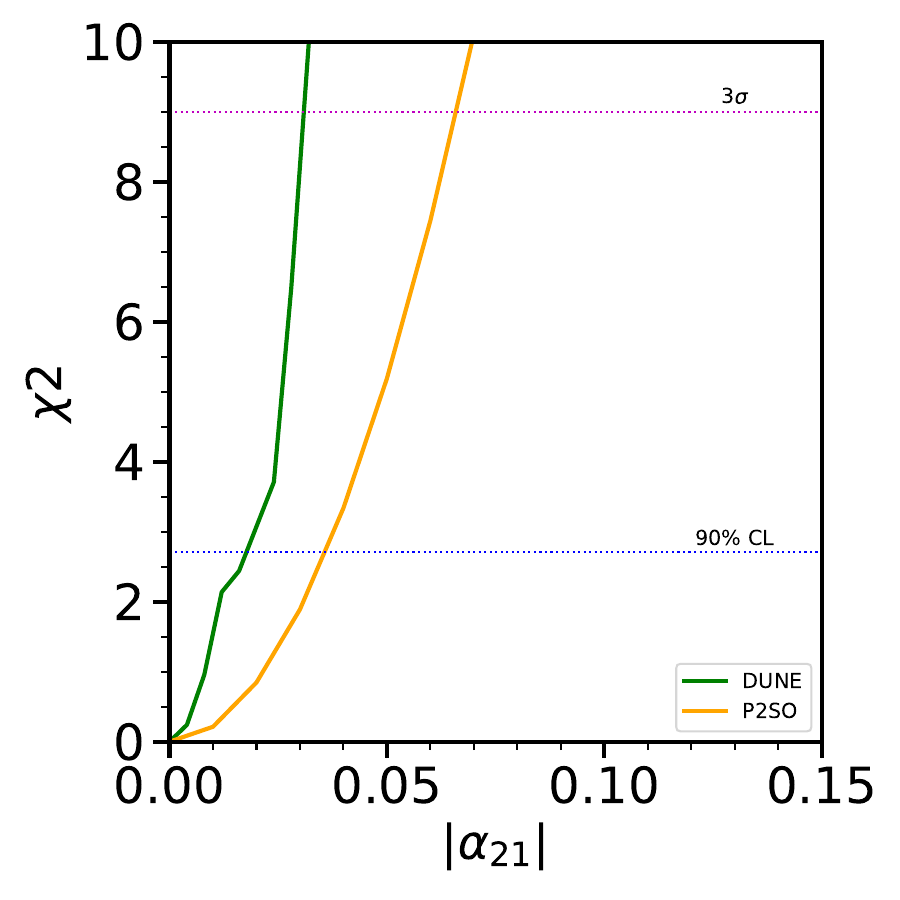} 
    \includegraphics[width=75mm, height=62mm]{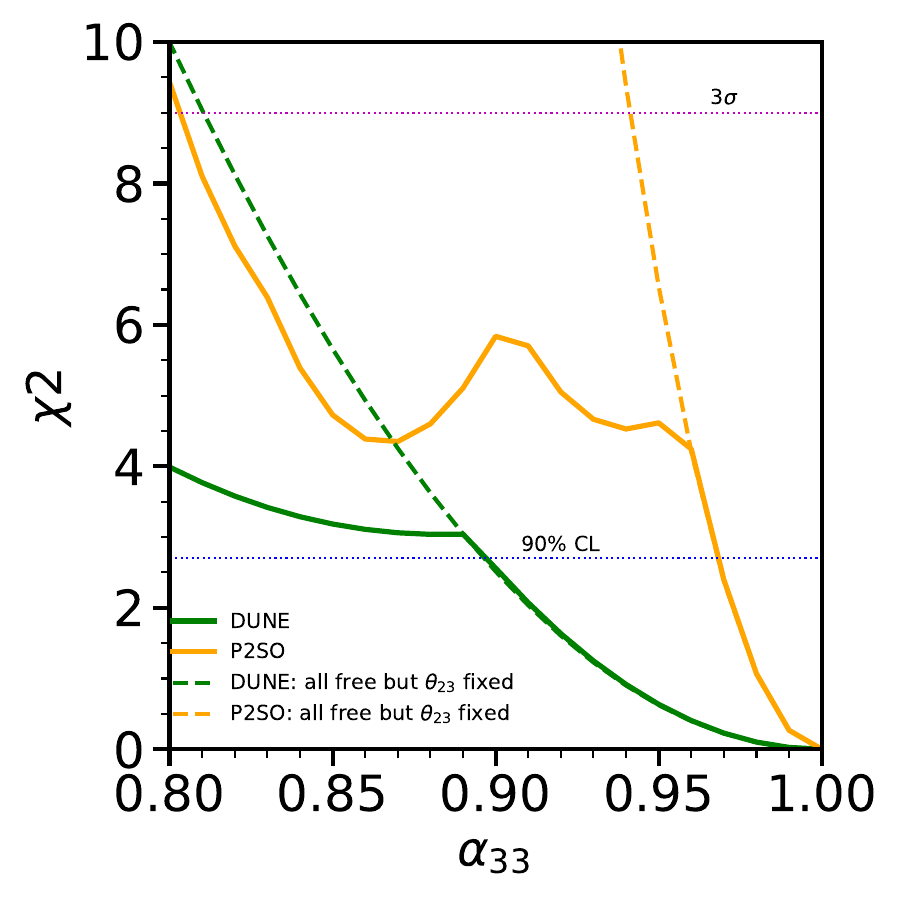}

     \caption{Sensitivity to NU parameters ($\alpha_{ij}$) for P2SO (orange) and DUNE (green). The horizontal lines  represent the  $3\sigma$ and $90\%$ C.L. The lower left panel illustrates the off-diagonal NU parameter while other three are diagonal NU parameters.}
    \label{boundplot}
    \end{center}
\end{figure}

First, we demonstrate the projected sensitivity to the NU parameters for P2SO experiment and compare it with that of DUNE in Fig. \ref{boundplot}. In the upper-left and upper-right panels, we show the constraints on $\alpha_{11}$ and $\alpha_{22}$ respectively, while the lower-left panel shows the bound on the off-diagonal parameter $\alpha_{21}$ for P2SO and DUNE. The lower-right panel presents the sensitivity to $\alpha_{33}$. In each panel, orange (green) curve represents the sensitivity results from P2SO (DUNE) experiment. Blue (purple) horizontal dashed lines represents the $90\%$ C.L. ($3\sigma$ C.L.). From Fig. \ref{boundplot}, we can see that, DUNE gives better  constraint on $\alpha_{11}$ and $\alpha_{21}$ whereas P2SO put stronger bound on $\alpha_{22}$ and $\alpha_{33}$. This is because, the parameter $\alpha_{33}$ does not enter the vacuum oscillation probabilities, but contributes in matter. Owing to its larger matter effects, P2SO is able to place a significantly stronger bound on $\alpha_{33}$ compared to DUNE, improving upon the current limits. DUNE, on the other hand, substantially strengthens the existing constraint on $\alpha_{11}$. Unlike other panels of Fig. \ref{boundplot}, the panel for $\alpha_{33}$ shows some unusual behavior. In the case of DUNE, the $\chi^2$  behaviour is smooth, with a kink appearing around $0.88-0.90$. However, the P2SO curve exhibits a pronounced dip at $0.85-0.87$, with a kink around $0.9-0.92$. To explain that, we analyze the impact of minimization on various oscillation parameters for both the experiments in the sensitivity curves. In the lower-right panel of  Fig. \ref{boundplot}, we present extra sensitivity curves of $\alpha_{33}$ with different sets of minimization of standard oscillation parameters given by the dashed curves. These curves represent the scenario in which all oscillation parameters are marginalized as described earlier, except for $\theta_{23}$, with the atmospheric mixing angle kept fixed at its true value. The unusual behavior of solid green and orange curves disappeared when we fix $\theta_{23}$. This indicates that the wavy nature in the sensitivity curves for $\alpha_{33}$ arises due to the degeneracy of $\alpha_{33}$ with the oscillation parameter $\theta_{23}$.


For generating Fig. \ref{boundplot}, we have taken one NU parameter at a time, however, nature can prefer all the NU parameters together. Thus, to have a realistic picture, ideally we need to vary all NU parameters together to get the bounds on these parameters. In Table \ref{1dbound}, we list the bounds of NU parameters in two different scenarios. We first mention the $90\%$ and $3 \sigma$ C.L. bounds of $\alpha_{11}, \alpha_{22}, \alpha_{33}$ and $\alpha_{21}$ by taking one NU parameter at a time (which we refer as ``one dof''), whereas in the next column, we give the bounds by varying all NU parameters simultaneously (which we refer as ``six dof''). The last column shows the present bound at $90\%$ C.L. with ``six dof'' configuration for each NU parameters. From the table, we can say that, DUNE is giving better result for $\alpha_{11}$ than present bound at $90\%$ C.L., while P2SO is giving better bound on $\alpha_{33}$ than the current bound. However, neither DUNE nor P2SO can put stronger bounds for $\alpha_{22}$ and $\alpha_{21}$ than the existing limits.


\begin{table}[h!]
\centering
\begin{tabular}{|c|c|c|c|c|c|c|c|c|c|}
\hline
\multirow{3}{*}{\textbf{NU Params}} & \multicolumn{4}{c|}{\textbf{one dof}} & \multicolumn{4}{c|}{\textbf{six dof}} & \multirow{3}{*}{\textbf{Current bounds \cite{Forero:2021azc} }} \\
\cline{2-9}
 & \multicolumn{2}{c|}{\textbf{DUNE}} & \multicolumn{2}{c|}{\textbf{P2SO}} & \multicolumn{2}{c|}{\textbf{DUNE}} & \multicolumn{2}{c|}{\textbf{P2SO}} & \\
\cline{2-9}
 & $90\% $ C.L. & $3 \sigma$ & $90\% $ C.L. &$ 3\sigma$ & $90\% $ C.L. & $3 \sigma$ & $90\% $ C.L. & $3 \sigma$ & ($90\%$ C.L., six dof) \\
\hline
$\alpha_{11}$ &0.984 & 0.971 &0.947 & 0.905 &\textbf{0.976} &0.956 &0.901 &0.800 & 0.969  \\ \hline
$\alpha_{22}$ &0.987 & 0.977&0.989 & 0.981& 0.986&0.974 &0.987 &0.976 & 0.995  \\ \hline
$\alpha_{33} $&0.896 & 0.564&0.967 & 0.803& 0.776&0.401 & \textbf{0.958}&0.792 & 0.890 \\ \hline
$\alpha_{21}$ &0.017 & 0.031&0.035 & 0.066& 0.028&0.052 & 0.049&0.110 & 0.013  \\
\hline
\end{tabular}
\caption{Constraints obtained on NU parameters from DUNE and P2SO with one and six degrees of freedom. Results are presented with both at $90\% $ and $3 \sigma$  C.L. }
    \label{1dbound}
\end{table}
\begin{figure}[h!]
\begin{center}
  \includegraphics[width=80mm, height=70mm]{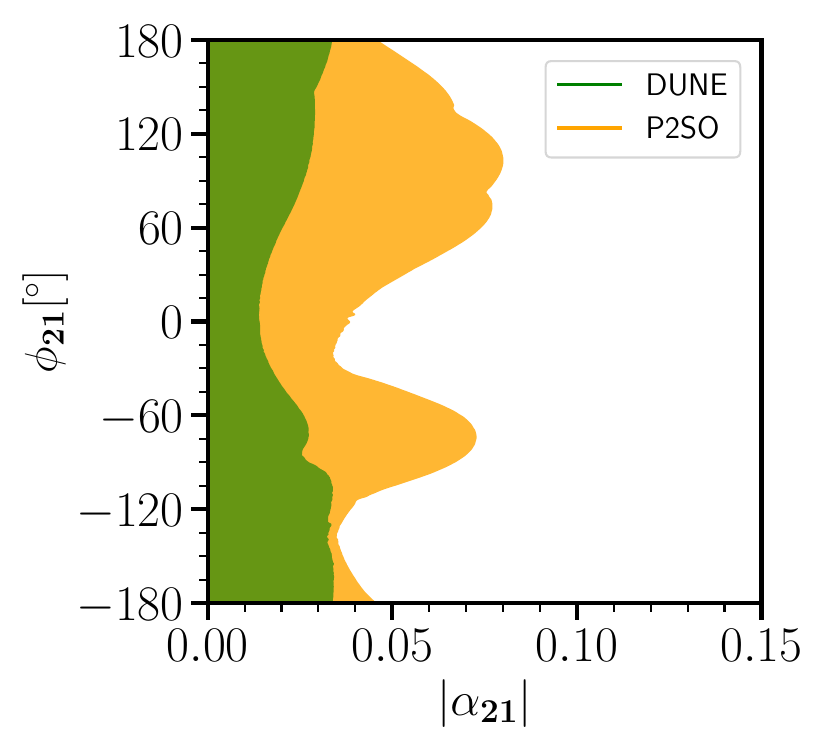}
    \caption{Allowed parameter space in the $|\alpha_{21}|$-$\phi_{21}$ plane, for DUNE (green) and P2SO (orange) experiments at $3 \sigma$ C.L.  }
    \label{alp-phi}
    \end{center}
\end{figure}

\subsection{Allowed parameter space between $|\alpha_{21}|$ and $\phi_{21}$}

Next, we study how the phase $\phi_{21}$ associated with the off-diagonal parameter $\alpha_{21}$ affects the bound on this parameter. Figure~\ref{alp-phi} shows the allowed parameter space of $|\alpha_{21}|-\phi_{21}$ at the $3\sigma$ confidence level assuming NU does not exist in Nature. The green (orange) contour is for  DUNE (P2SO) experiment. We see that the sensitivity to the off-diagonal parameter is strongly influenced by the value of the corresponding phase. This dependence is more pronounced in the case of P2SO compared to DUNE.  For P2SO, the sensitivity is notably weak around $\phi_{21} = 100^\circ$ and   $-75^\circ$, leading to poor upper bounds. In contrast, when the phase is set to $0$, the bounds become significantly stronger for both experiments.

\begin{figure}[h!]
\begin{center}
    \includegraphics[width=75mm, height=62mm]{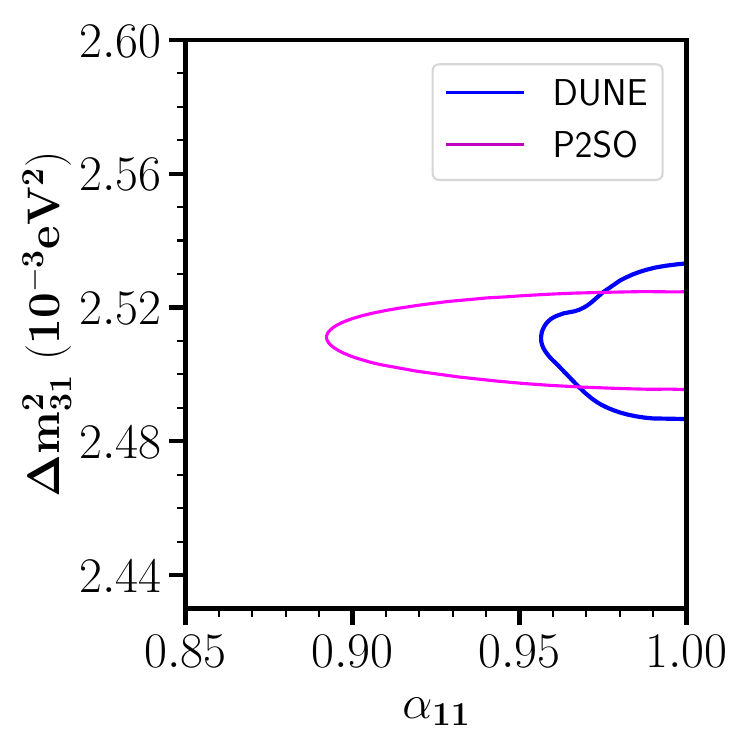}
    \includegraphics[width=75mm, height=62mm]{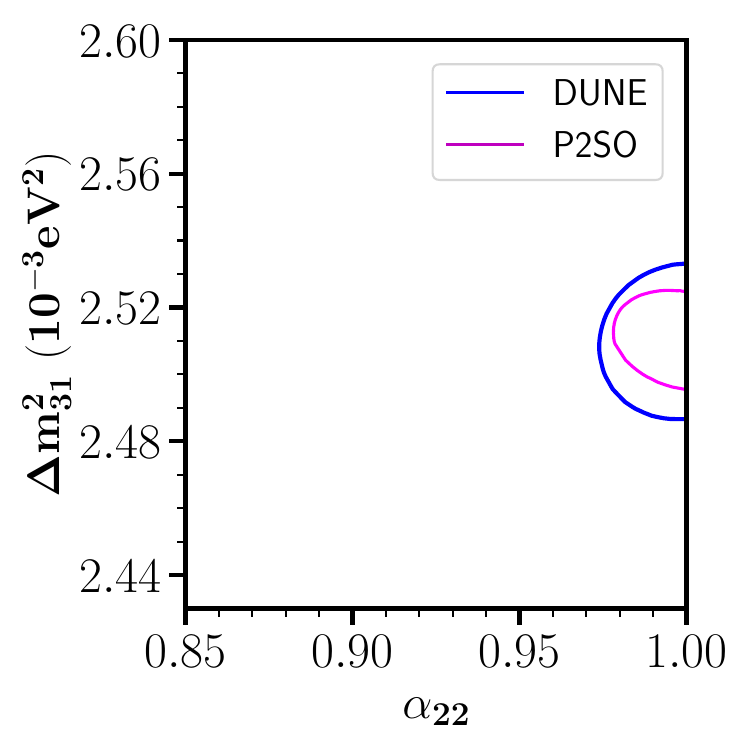}  \\
    \includegraphics[width=75mm, height=62mm]{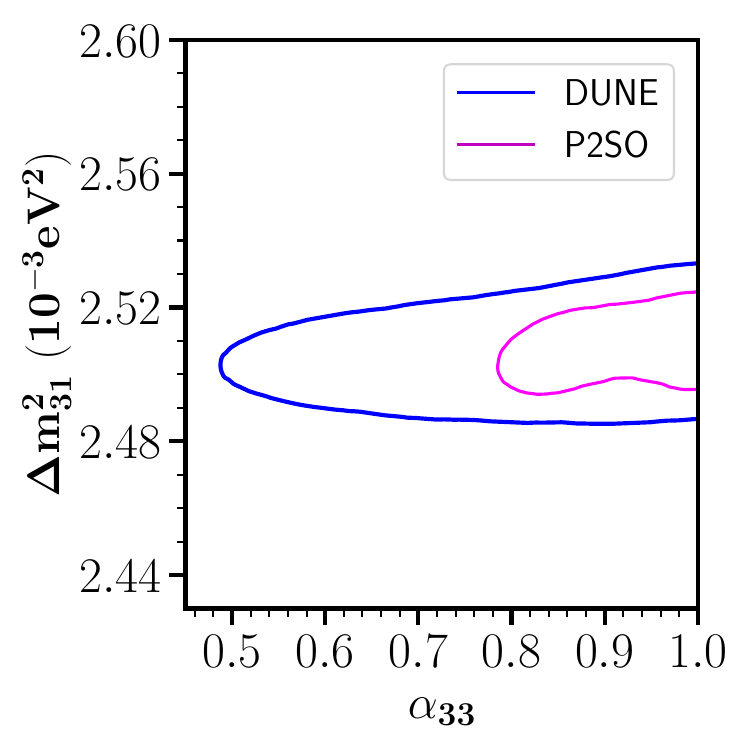}
    \includegraphics[width=75mm, height=62mm]{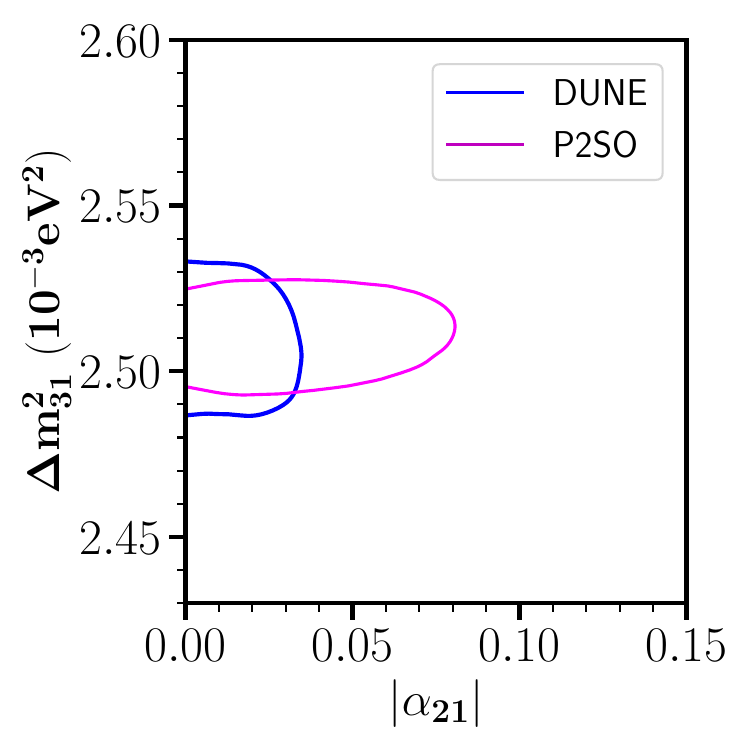}  
   
    \caption{ Allowed parameter space between $\alpha_{ij}-\Delta m^2_{31}$ at $3\sigma$ C.L. In each panel, magenta (blue) contour shows the parameter space for P2SO (DUNE) experiments at $3 \sigma$ C.L.}
    \label{alpldm}
\end{center}
\end{figure}

\subsection{Allowed parameter space between $\alpha_{ij}$ and $\Delta m^2_{31}$}

In this subsection, we show the correlation of atmospheric mass splitting ($\Delta m_{31}^2$) with NU parameters, assuming NU does not exist in nature. Figure~\ref{alpldm} shows how $\Delta m_{31}^2$ depends on the NU parameters; $\alpha_{11}$ (upper-left), $\alpha_{22}$ (upper-right), $\alpha_{33}$ (lower-left), and $\alpha_{21}$ (lower-right). In each panel, blue and magenta contours represent allowed parameter space for the DUNE and P2SO experiments, respectively, with all contours plotted at the $3\sigma$ confidence level.  The $y$-axis corresponds to the currently allowed $3\sigma$ range of $\Delta m_{31}^{2}$.  For the panels with $\alpha_{ii} ~(i=1,2,3)$, the rightmost point represents the standard case, while for $\alpha_{21}$, the zero value corresponds to standard oscillation scenario. 
 The closed nature of these contours demonstrates that marginalizing over $\Delta m_{31}^{2}$ within its present global $3\sigma$ range is sufficient to constrain the NU parameters precisely.

\subsection{Effect on Mass hierarchy}
\label{EMH}
\begin{figure}[h!]
\begin{center}
    \includegraphics[width=75mm, height=62mm]{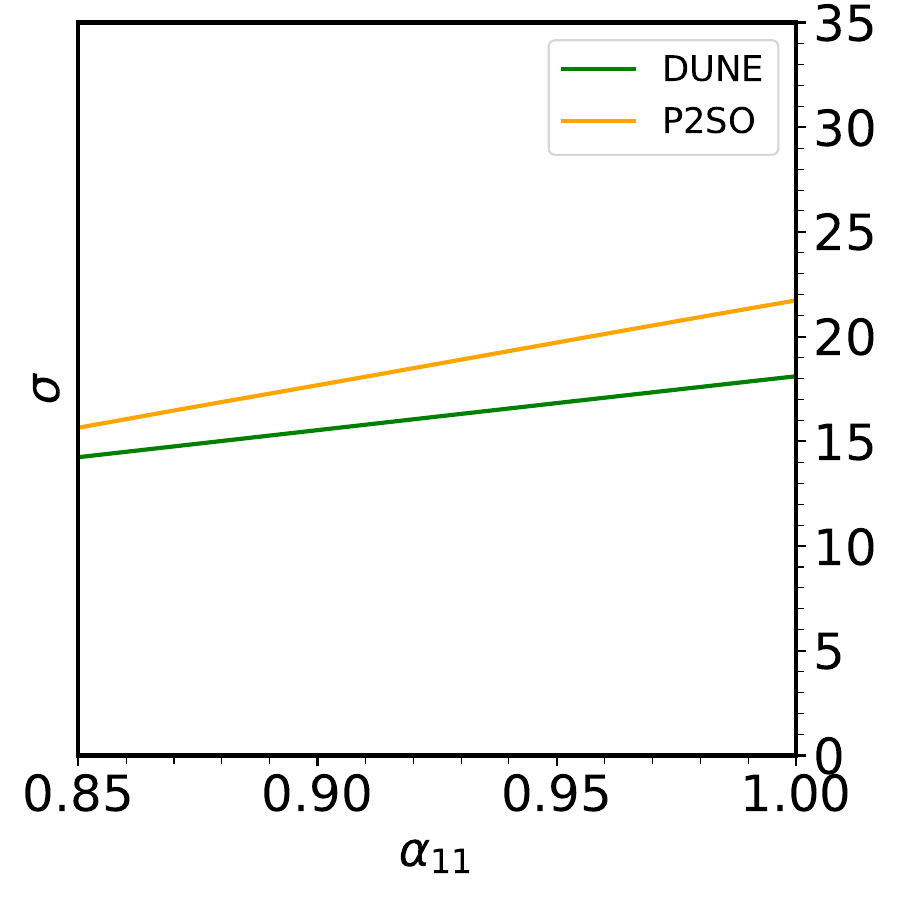}
    \includegraphics[width=75mm, height=62mm]{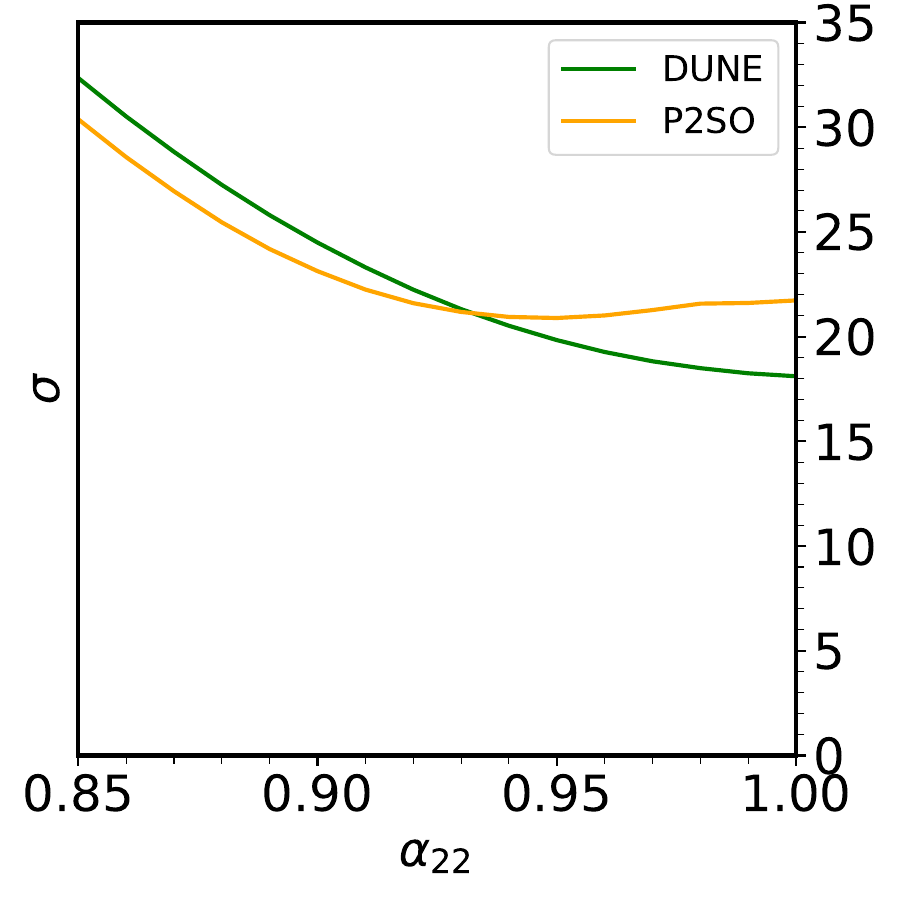}  
      \\
    \includegraphics[width=75mm, height=62mm]{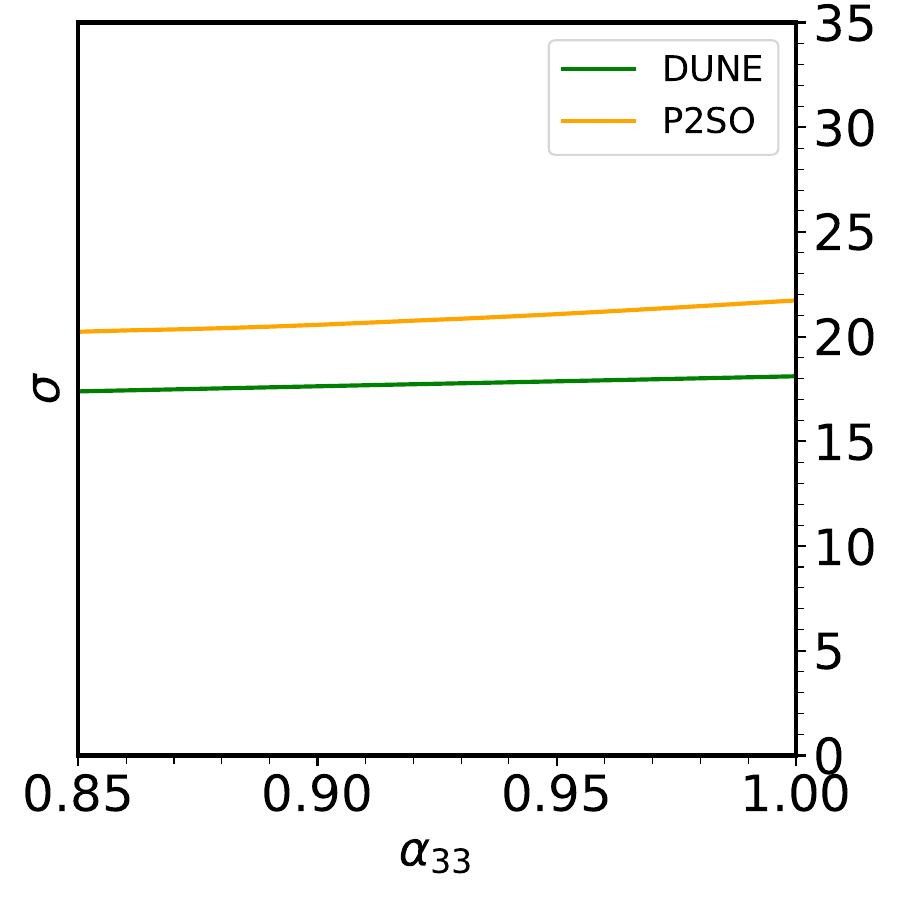}
    \includegraphics[width=75mm, height=62mm]{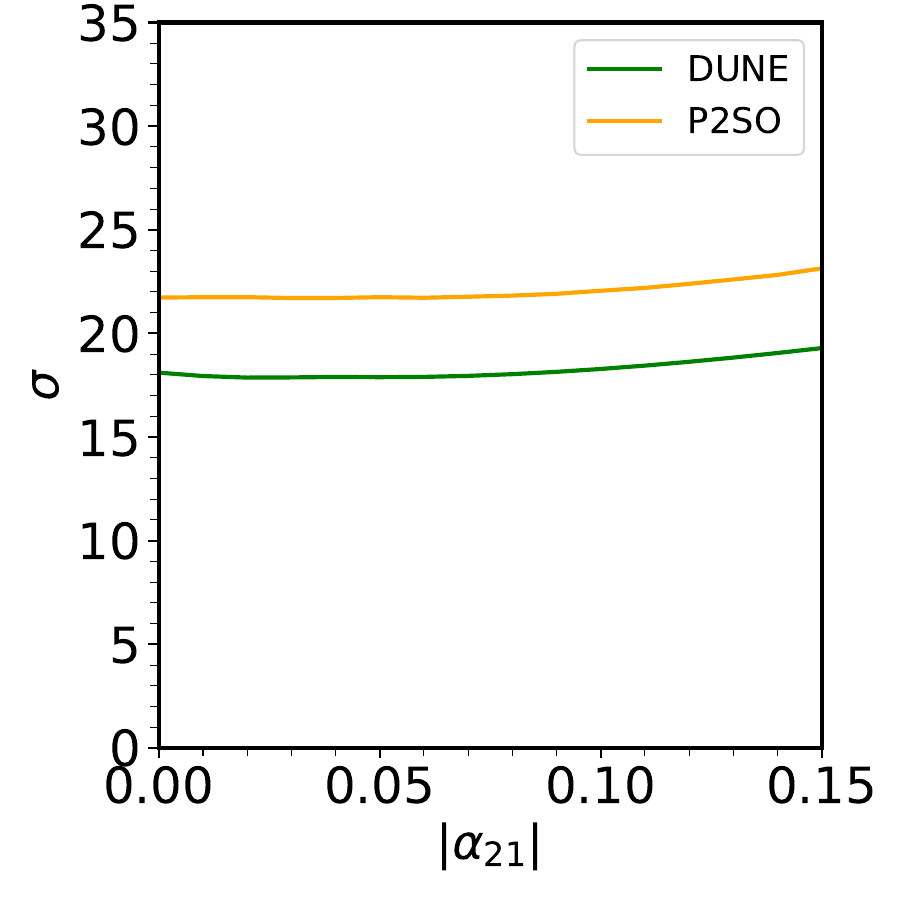}  
   
    \caption{Mass hierarchy sensitivity in the presence of NU parameters ($\alpha_{ij}$) for P2SO (orange) and DUNE (green).}
    \label{mhsen}
    \end{center}
\end{figure}

\begin{figure}
    \centering
    \includegraphics[width=75mm, height=62mm]{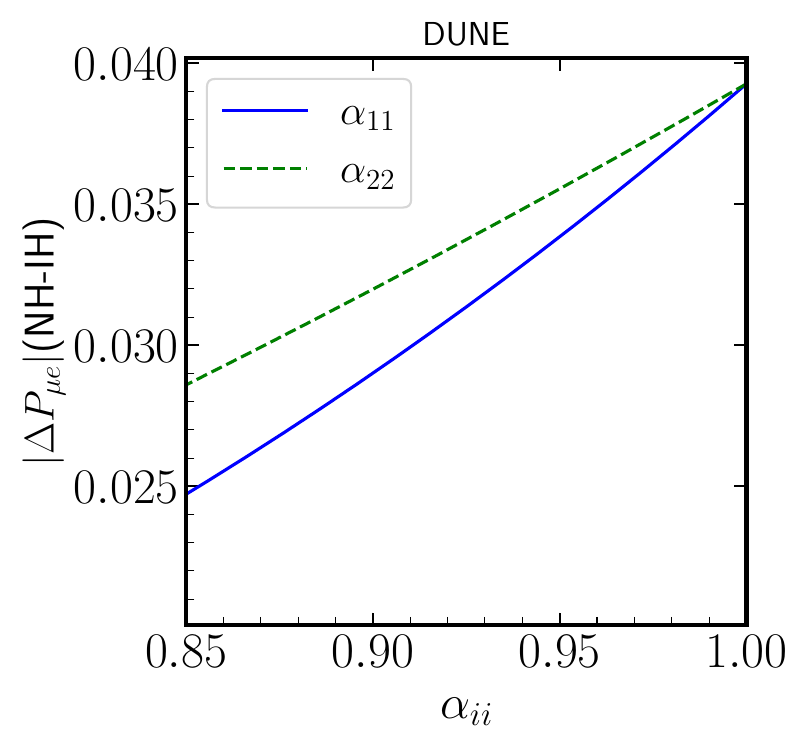}
    \includegraphics[width=75mm, height=62mm]{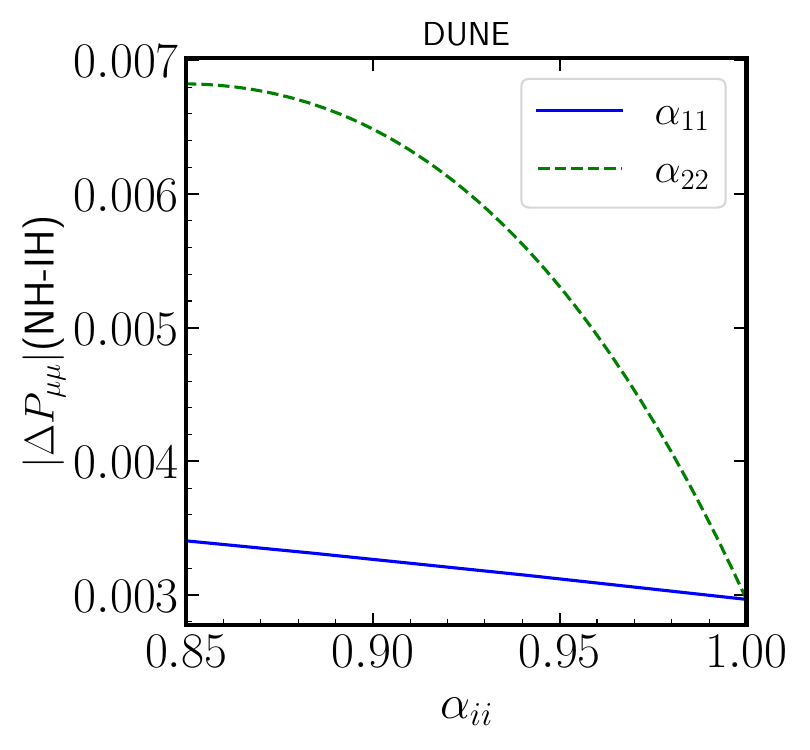}  
    
    \includegraphics[width=75mm, height=62mm]{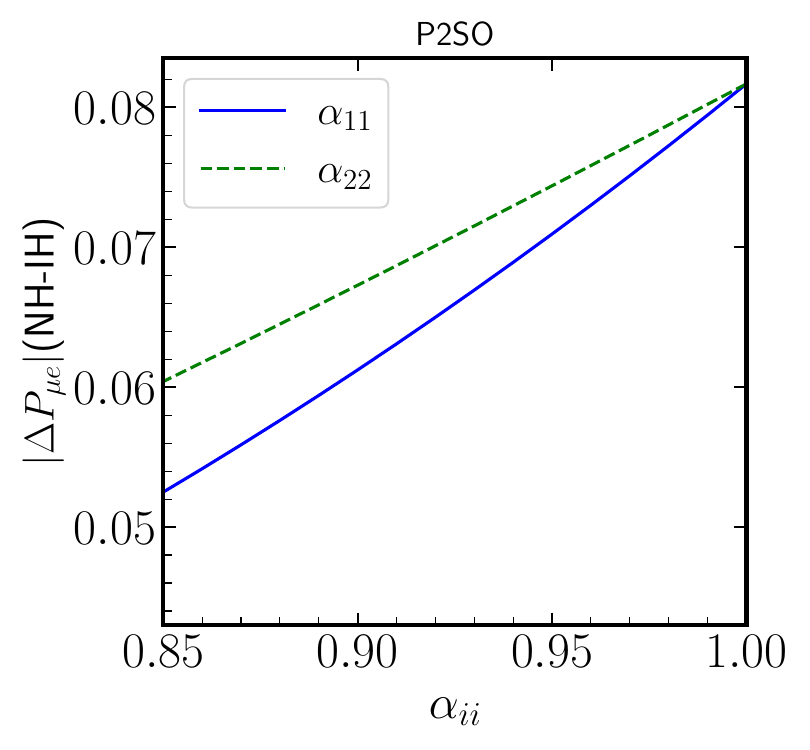}
    \includegraphics[width=75mm, height=62mm]{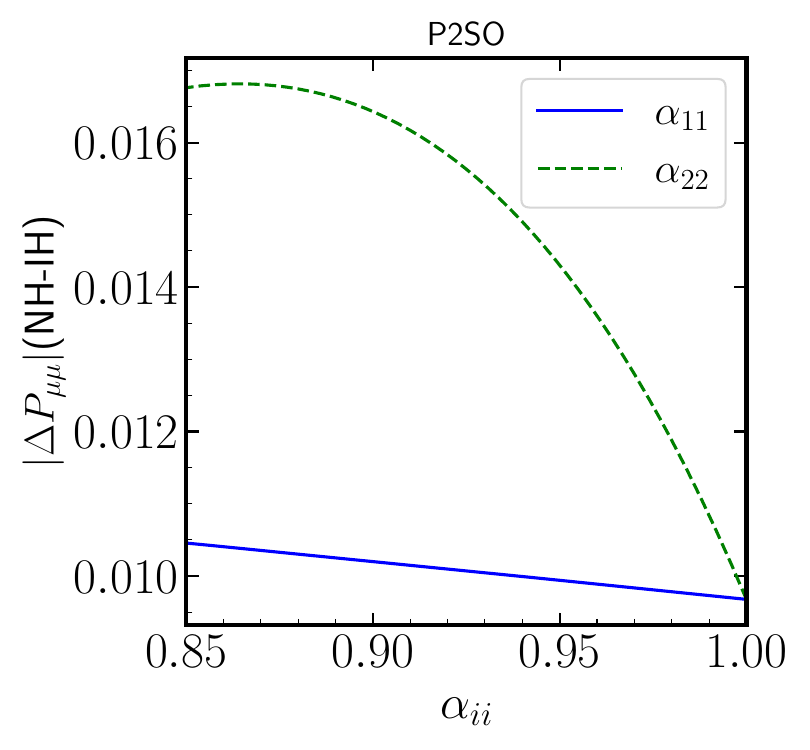}  

    \caption{Appearance (left) and disappearance (right) probability difference (NH-IH)  as a function of $\alpha_{ii}$ 
 for DUNE and P2SO.}
    \label{fig:NH-IH}
\end{figure}
Mass hierarchy sensitivity of any experiment is the capability of excluding wrong hierarchy from correct one. In this section, we show the variation of mass hierarchy sensitivity with respect to the true values of the NU parameters for the two experiments, P2SO and DUNE. The $y$ axis of Fig.~\ref{mhsen} shows the value of mass hierarchy sensitivity, $\sigma (= \sqrt{\chi^2)}$ whereas $x$ axis shows the variation of NU parameters; $\alpha_{11}$ (upper-left), $\alpha_{22}$ (upper-right), $\alpha_{33}$ (lower-left) and $\alpha_{21}$ (lower-right). The NU parameters are fixed in test\footnote{Note that in Ref.~\cite{Kaur:2021rau}, while determining the physics sensitivities, standard scenario was considered in the true spectrum of the $\chi^2$. They considered NU in the test spectrum and showed their sensitivity as function of $\delta_{\rm CP}$ for different ratios of neutrino to antineutrino runtime.}. In the lower-right panel, the true value of the complex phase $\phi_{21}$ is taken as $ 0$. For generating each plot, we consider normal hierarchy (NH) in the true scenario and inverted hierarchy (IH)  in the test scenario. We follow the minimization procedure as mentioned in Section \ref{res}. In each panel, orange (green) curve shows the mass hierarchy sensitivity variation as a function of NU parameters for P2SO (DUNE) experiment. For $\alpha_{11}$, we observe a significant reduction in the mass hierarchy sensitivity as the parameter deviates from its standard value ($\alpha_{11}=1$) in both experiments. For P2SO, the sensitivity remains almost flat in the deviation of $\alpha_{22}$ from its standard value ($\alpha_{22}=1$) up to 0.93 and increases gradually beyond this value, whereas for DUNE, it shows a steady rise starting from the standard scenario. On the other hand, the sensitivity is nearly unchanged for all values of $\alpha_{33}$. It should be emphasized that, there is no strong dependency of $\alpha_{33}$ on both $P_{\mu e}$ and $P_{\mu \mu}$ in vacuum. The small change in $\sigma$ due to $\alpha_{33}$ comes only from matter effects. In presence of $\alpha_{21}$, the mass hierarchy sensitivity slightly increases for both the experiments.

It can be noted that, the behavior of mass hierarchy sensitivity is contradictory for $\alpha_{11}$ and $\alpha_{22}$ panels, \textit{i.e.}, decreasing for $\alpha_{11}$ and increasing for $\alpha_{22}$. We try to understand this in Fig.~\ref{fig:NH-IH}. This figure presents the variation in probability difference ($|\Delta P_{\alpha \beta}| =P_{\alpha\beta}$ (NH)$-P_{\alpha\beta}$ (IH)) as a function of $\alpha_{ii}$, with the neutrino energy fixed at 2.5 GeV for DUNE and 5 GeV for P2SO. The left column displays the appearance probability, while the right column corresponds to the disappearance channel. The top and bottom rows represent the DUNE and P2SO experiments, respectively. In each panel, blue solid (green dashed) curve represents the change in probability with respect to $\alpha_{11}$ ($\alpha_{22})$. From the figure, we can see that, $|\Delta P_{\mu e}|$ decreases in presence of NU parameters. This is true for both the parameters, $\alpha_{11}$ and $\alpha_{22}$. For the case of $|\Delta P_{\mu \mu}|$, the probability difference increases in presence of both the NU parameters. However, the rate of increase for $\alpha_{22}$ is significantly higher than for $\alpha_{11}$. Therefore, the hierarchy sensitivity of $\alpha_{22}$ gets dominated by the disappearance channel and we observe an overall increase in the hierarchy sensitivity. This observation is similar for both the experiments.

The above can be also supported via analytic expressions of appearance and disappearance probabilities mentioned in Eqs. \ref{eq:P_mue} and \ref{eq:P_mumu} \footnote{Although Eqs. \ref{eq:P_mue} and \ref{eq:P_mumu} are derived in vacuum, they can still be used to understand the impact of NU parameters on mass-hierarchy sensitivity, as they exhibit patterns similar to those arising from matter effects.}. From the expressions, it is evident that the appearance probability depends on $\alpha_{11}$, $\alpha_{22}$, and $\alpha_{21}$, while the disappearance probability depends only on $\alpha_{22}$ and $\alpha_{21}$. Consequently, $\alpha_{11}$ affects only the appearance channel, whereas $\alpha_{22}$ can have a significant impact on both probability channels.

\subsection{Effect on Octant Sensitivity}

\begin{figure}[h!]
\begin{center}
    \includegraphics[width=75mm, height=62mm]{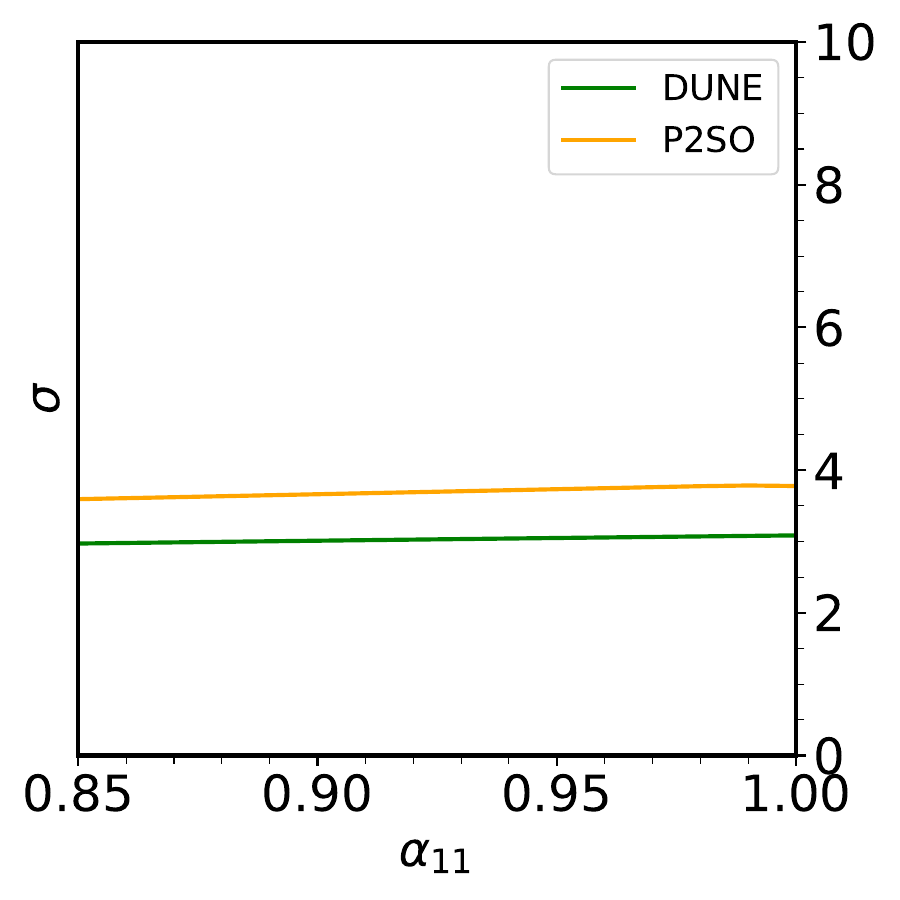}
    \includegraphics[width=75mm, height=62mm]{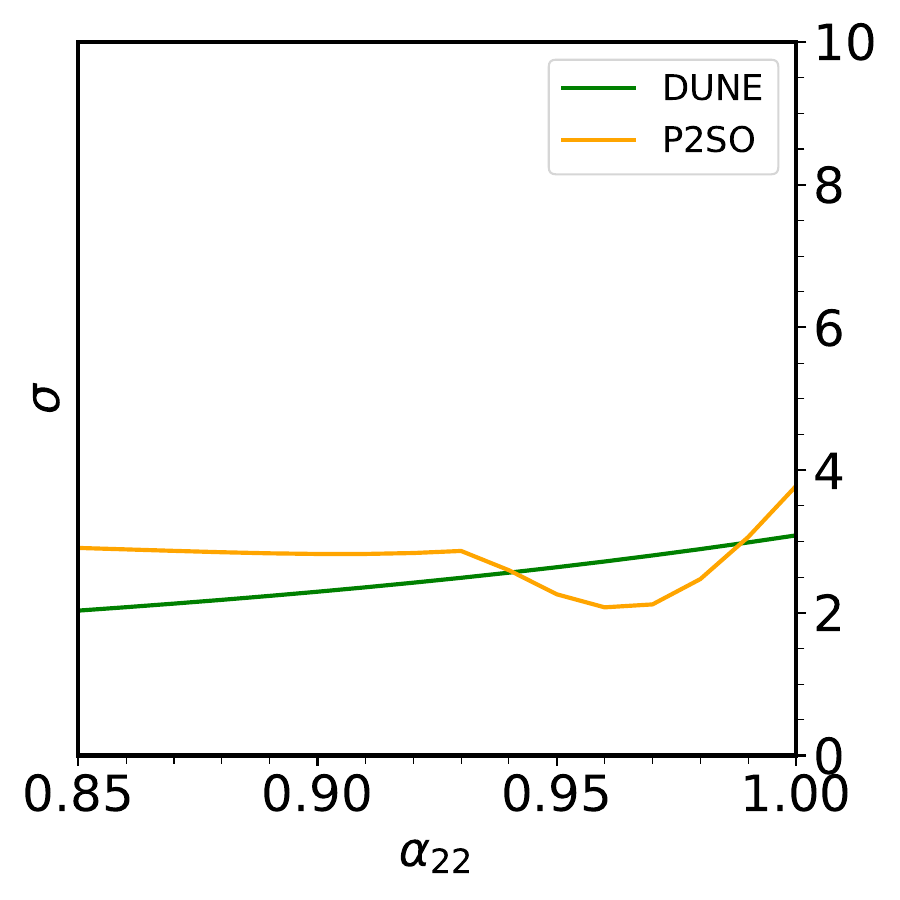}    \\
    \includegraphics[width=75mm, height=62mm]{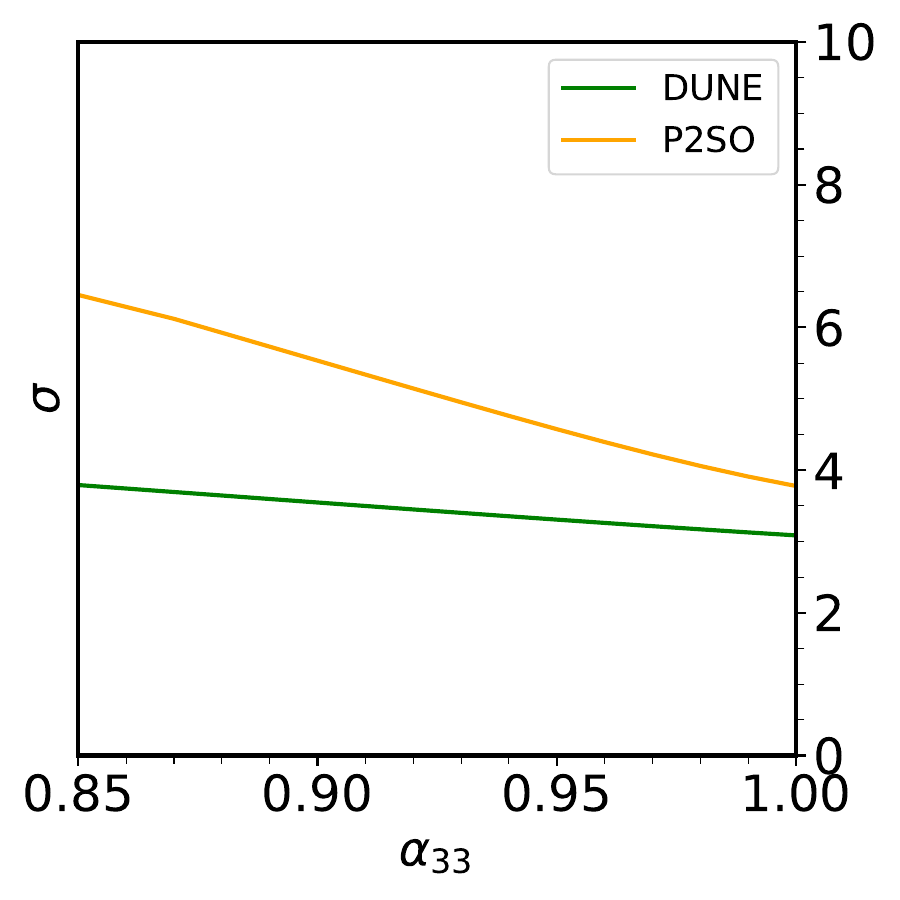}
    \includegraphics[width=75mm, height=62mm]{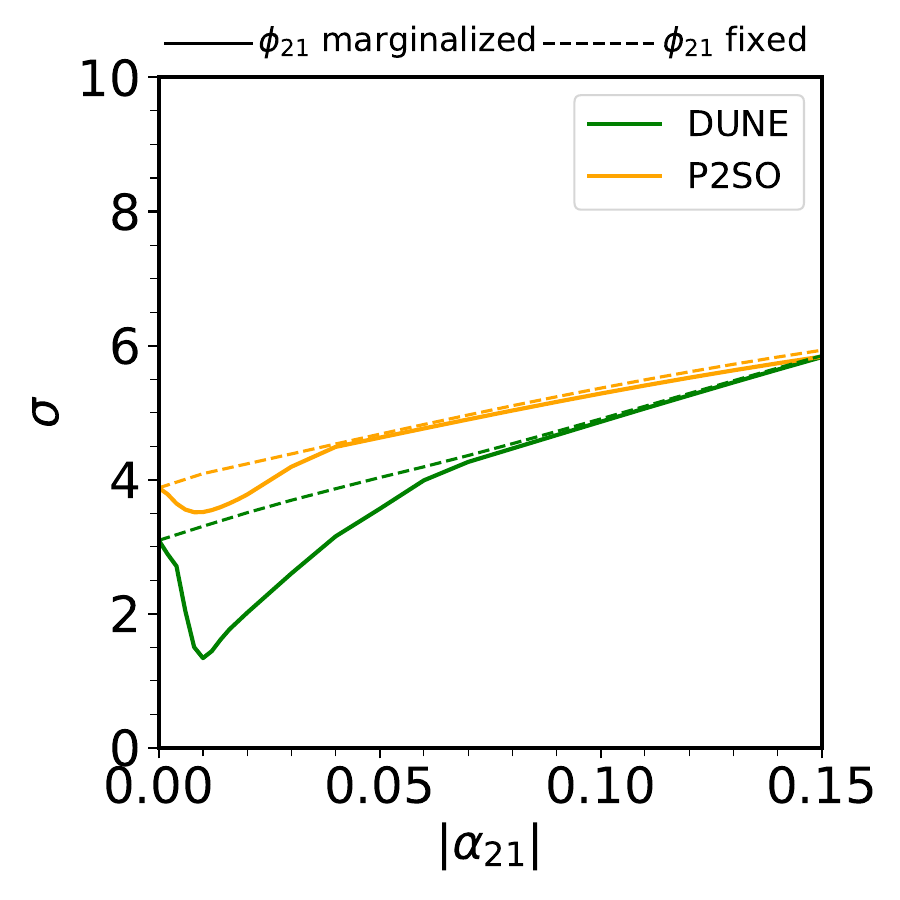}  
     \caption{Octant sensitivity in the presence of NU parameters ($\alpha_{ij}$) for P2SO (orange) and DUNE (green).  }
    \label{octsen}
    \end{center}
\end{figure}

\begin{figure}[h!]
\begin{center}
    \includegraphics[width=75mm, height=62mm]{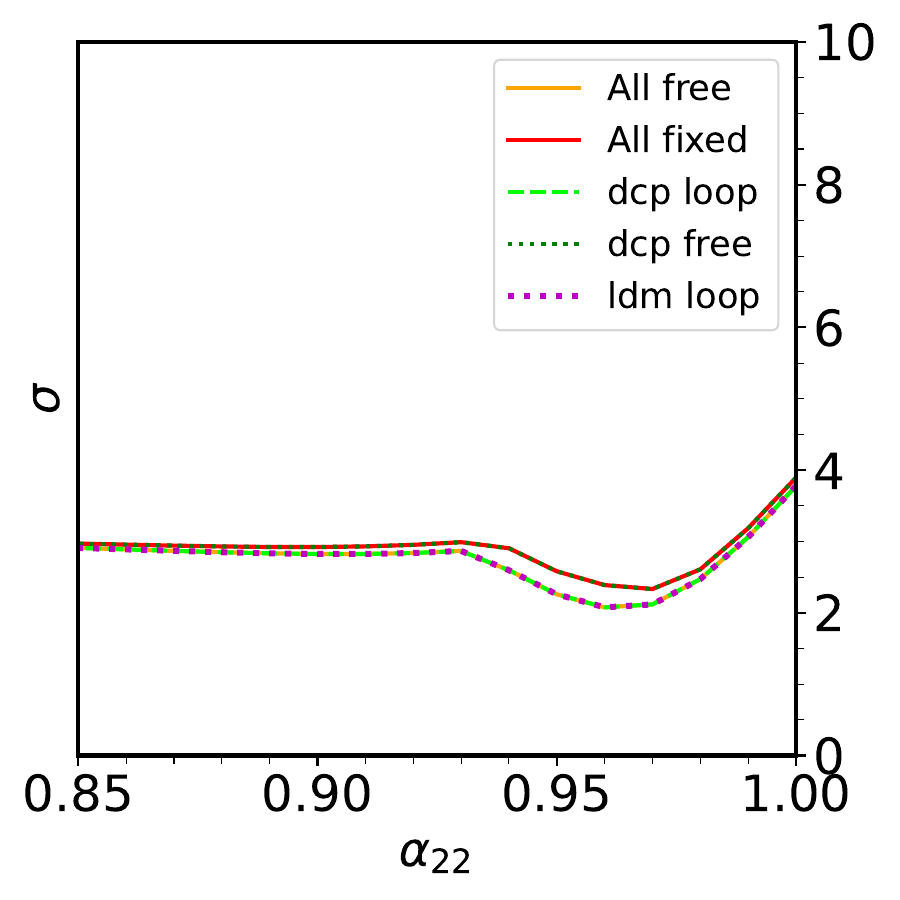} 
    \includegraphics[width=75mm, height=62mm]{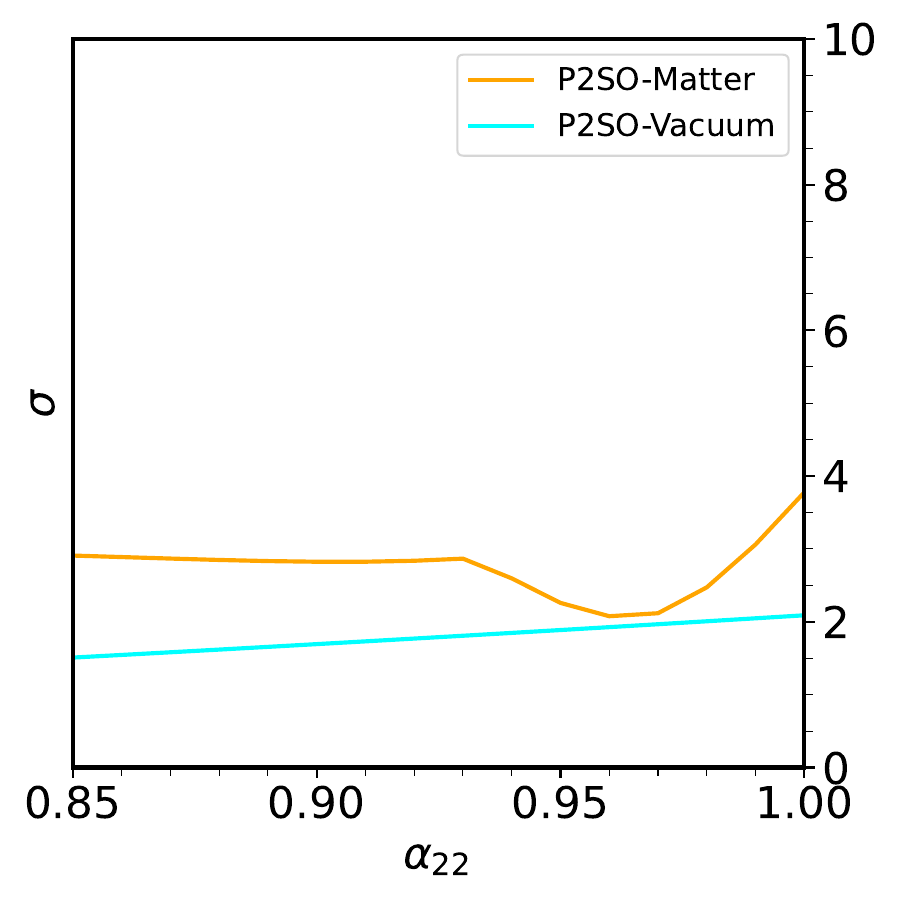}\\
    \includegraphics[width=75mm, height=62mm]{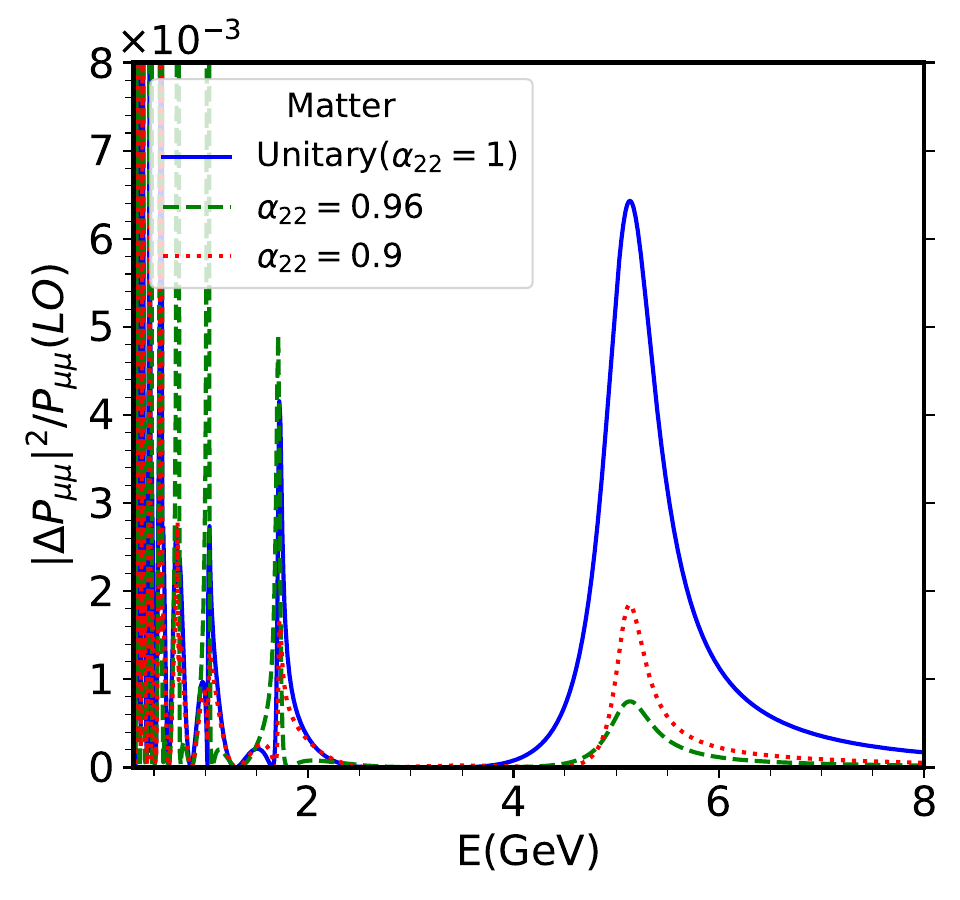}  
    \includegraphics[width=75mm, height=62mm]{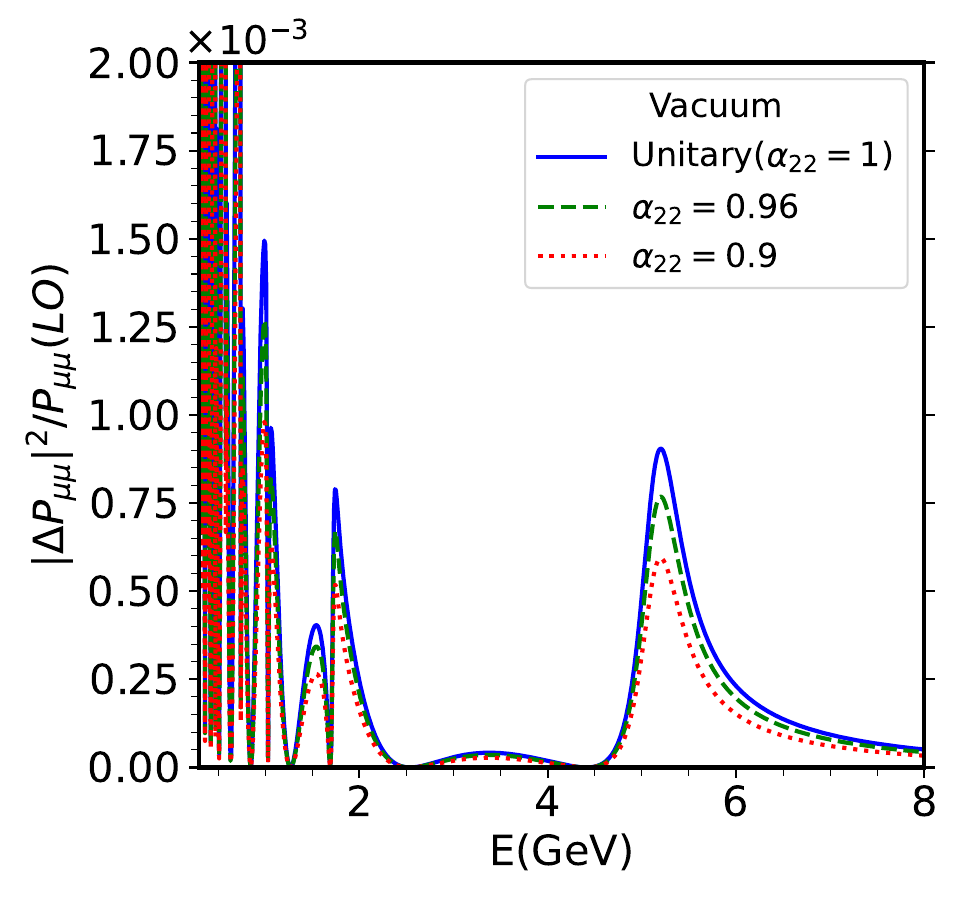}  
    \caption{Octant sensitivity for P2SO with the NU parameter $\alpha_{22}$. Upper panels: effects of parameter marginalization (left) and matter effects (right). Lower panels: relative probabilities with fixed NU values in matter (left) and in vacuum (right).}
    \label{octsenexplain}
    \end{center}
\end{figure}

In this subsection, we show the octant sensitivity as a function of true values of the NU parameters, $\alpha_{ij}$. Octant sensitivity of any experiment is the ability to exclude the wrong octant from correct one. In our simulation, we take the lower octant (LO) in the true scenario and the higher octant (HO) in the test scenario. The sensitivity is presented as a function of $\alpha_{ij}$ in Fig.~\ref{octsen}. Similar to mass hierarchy, the $y$ axis of Fig. \ref{octsen} shows the octant sensitivity where $x$ represents the variation of NU parameters; $\alpha_{11}$ (upper-left), $\alpha_{22}$ (upper-right), $\alpha_{33}$ (lower-left) and $\alpha_{21}$ (lower-right). The NU parameters are fixed in the test. In the lower-right panel, the true value of the complex phase $\phi_{21}$ is taken as $0$. The minimization procedure is similar as mentioned in Section \ref{res}. In each panel, orange (green) curve represents the octant sensitivity in variation of NU parameters for P2SO (DUNE) experimental setup. From the figure, we can see, there is very minimal change in the sensitivity with respect to the variation of $\alpha_{11}$ parameter.  
However, for $\alpha_{22}$ there is a significant decrease in octant sensitivity value in deviating the value of $\alpha_{22}$ for DUNE, while for P2SO there is a small dip around $\alpha_{22}=0.97$. In lower-left panel, the sensitivity increases sharply for both P2SO and DUNE for the NU parameter $\alpha_{33}$. Bottom right panel shows the octant sensitivity in the presence of the complex parameter $\alpha_{21}$, illustrating the impact of both its magnitude and associated phase $\phi_{21}$. The  solid curves corresponding to the case where $\phi_{21}$ is marginalized over full range, and dashed curves representing fixed $\phi_{21}$ ($\phi_{21}=0$) value. The panel shows that for both the experiments, octant sensitivity increases in presence of $\alpha_{21}$ when the complex phase is fixed ($\phi_{21}=0$). However, marginalising over the NU phase leads to a degeneracy region near $\alpha_{22}=0.01$, and this behaviour is observed in both experiments. It demonstrates that the complex phase of $\phi_{21}$ plays a major role in determining octant sensitivity, and underlines the importance of off-diagonal NU parameter in precision measurements of $\theta_{23}$ parameter.

Let us now try to understand the origin of the dip in P2SO for $\alpha_{22}$. To check if it appears due to degeneracy of the standard oscillation parameters with the NU parameter $\alpha_{22}$, in the upper-left panel of Fig.~\ref{octsenexplain}, we study the effect of marginalization of the standard oscillation parameters. In this panel we fix all oscillation parameters and then allow one parameter to vary at a time to search for possible degeneracies. Our analysis shows that none of the parameters account for the dip, suggesting that this peculiar behavior likely originates due to matter effect. In upper-right panel of Fig.~\ref{octsenexplain}, we plot octant sensitivity in matter (orange) and in vacuum (cyan). From the panel, it is clearly visible that, the dip is coming due to matter effect in P2SO. To understand this further, we perform a probability level analysis, as illustrated in lower panels of Fig.~\ref{octsenexplain}. Lower-left (lower-right) panel shows the ratio of square of $| \Delta P_{\mu \mu}|$ (where $|\Delta P_{\mu \mu}| = P_{\mu \mu} ~(\rm{HO}) - P_{\mu \mu} ~(\rm{LO})$) to $P_{\mu \mu} ~(\rm{LO})$ as a function of energy in presence of matter (vacuum) for three different $\alpha_{22}$ values; blue curve is with the standard case ($\alpha_{22}=1$), green dashed is for the value of $\alpha_{22}=0.96$ where the dip occurs, red dotted curve is with another value of $\alpha_{22}$ as 0.9, where no dip is present. Octant sensitivity is proportional to $|\Delta P_{\mu \mu}|^2$, thus the result of lower panel of Fig. \ref{octsenexplain} directly reflects the behavior of octant sensitivity presented in Fig.~\ref{octsen}~\footnote{It is to note that, the octant sensitivity also depends on appearance probability, thus similar plots can be done with $P_{\mu e}$ channel, however, we have seen that the effect of $\alpha_{22}$ on $P_{\mu e}$ is small compared to $P_{\mu \mu}$ channel.}. From the lower panels of Fig. \ref{octsenexplain}, we see that around 5 GeV, the green curve lies below the red curve for matter but is the opposite in vacuum. This explains the dip at $\alpha_{22}=0.96$ for P2SO in matter.


\subsection{CP violation sensitivity}

\begin{figure}[h!]
\begin{center}
    \includegraphics[width=75mm, height=62mm]{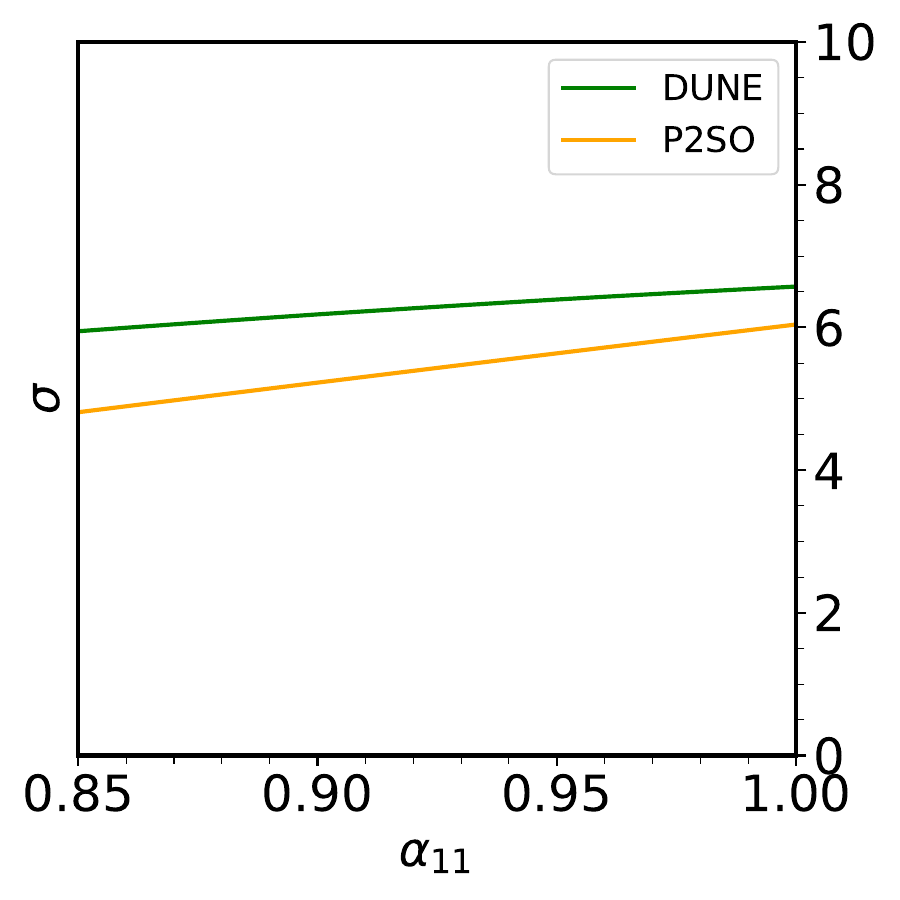}
    \includegraphics[width=75mm, height=62mm]{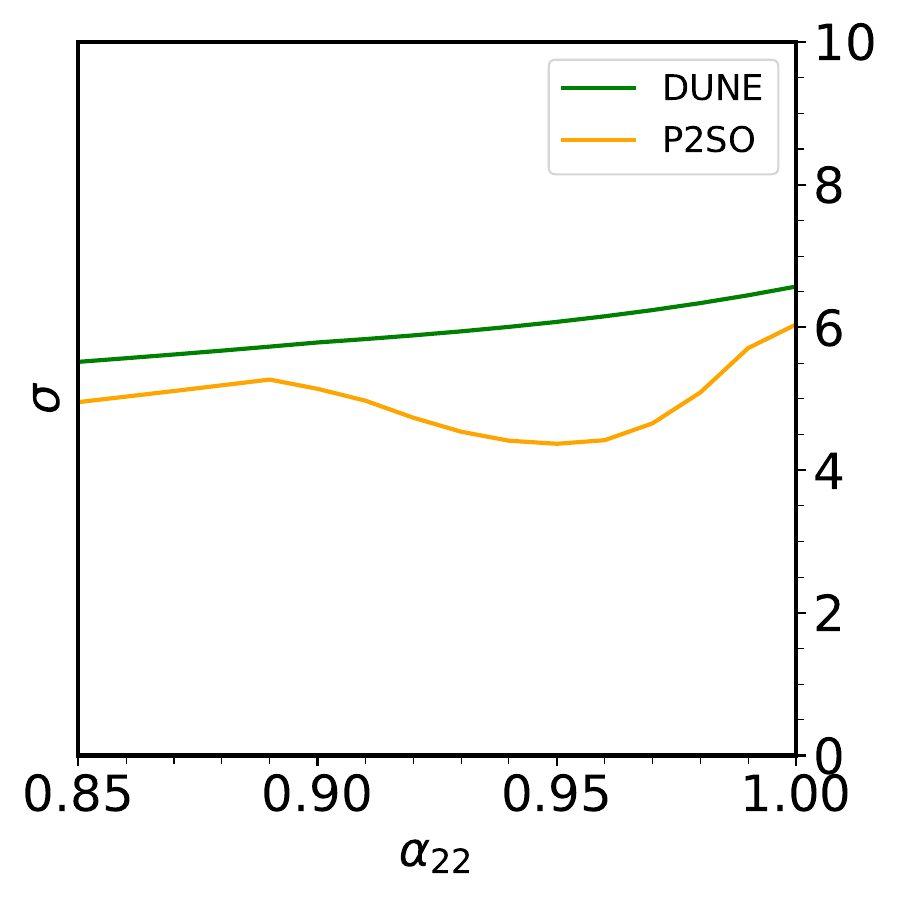}  \\
    \includegraphics[width=75mm, height=62mm]{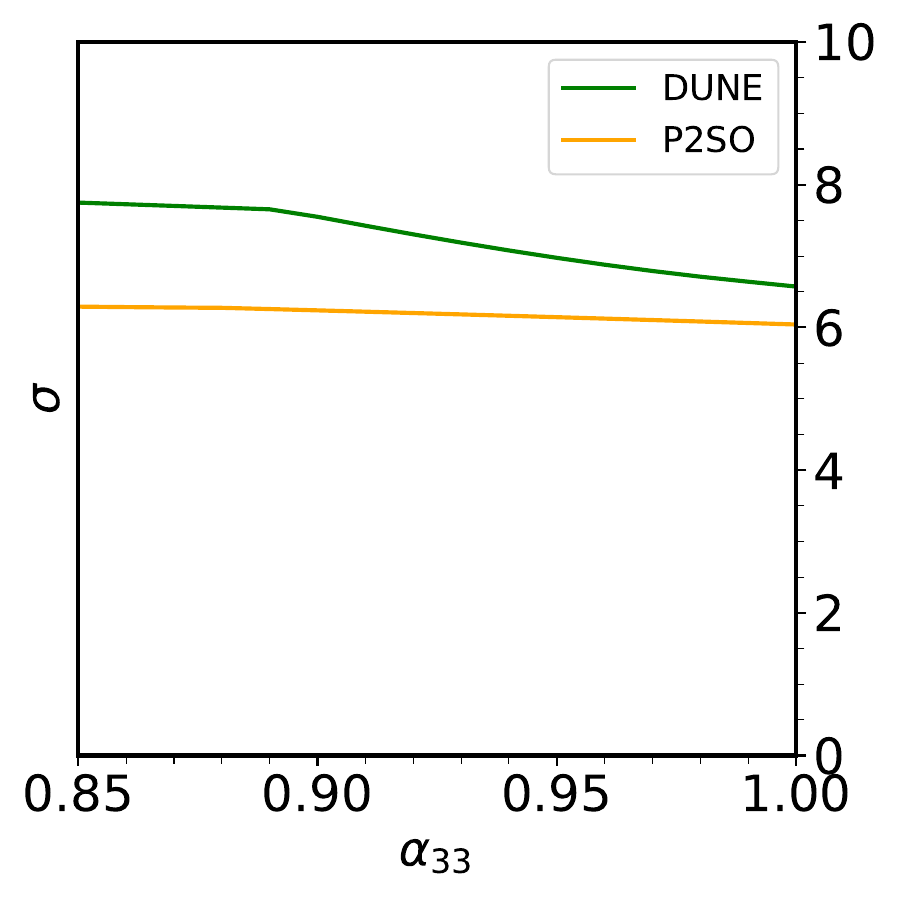}
    \includegraphics[width=75mm, height=62mm]{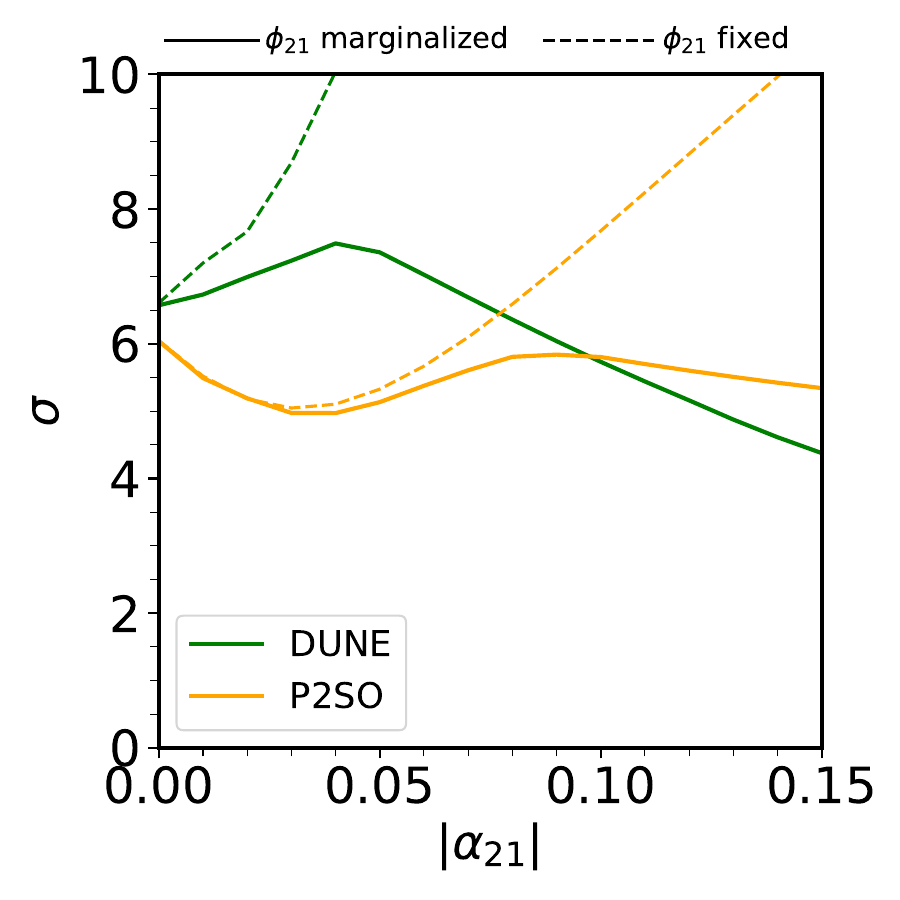}  
   
    \caption{CP violation sensitivity in the presence of NU parameters. P2SO (DUNE) and DUNE (green). }
    \label{cpsen}
    \end{center}
\end{figure}
The measurement of the leptonic CP-violating (CPV) phase, $\delta_{\text{CP}}$, remains one of the most challenging and least constrained parameter of neutrino oscillation physics. The CPV sensitivity represents the ability of an experiment to rule out CP-conserving values of $\delta_{\text{CP}}$ i.e.,  $ 0$ or $180^\circ$. In Fig.~\ref{cpsen}, we show the CPV sensitivity as a function of NU parameters; $\alpha_{11}$ (upper-left), $\alpha_{22}$ (upper-right), $\alpha_{33}$ (lower-left) and $\alpha_{21}$ (lower-right). In our simulation, we have taken $\delta_{CP}^{\rm true}= -90^{\circ}$. The marginalization procedure for all other oscillation parameters are same as mentioned in Section \ref{res}.  The orange curves correspond to the results from P2SO experiment, while the green curves represent DUNE. From the figure, we see in the presence of the NU parameters $\alpha_{11}$ and $\alpha_{22}$, the sensitivity decreases for both experiments, while for the parameter $\alpha_{33}$, the sensitivity increases slightly in presence of NU parameter. In the lower-right panel, in presence of $\alpha_{21}$, the behaviour of the CPV sensitivity curves differ between the two experiments. 

It is worth noting that the CP-violation sensitivity for the P2SO experiment exhibits a dip around $\alpha_{22}=0.96$, a behavior that closely resembles the octant sensitivity pattern observed in the upper-right panel of Fig.~\ref{octsen}. Hence, the dip in CPV sensitivity for P2SO with $\alpha_{22}$ is caused by matter effects, in a way similar to the octant sensitivity behavior. 

Further, in the lower-right panel,  sensitivity decreases for both experiments after $|\alpha_{21}|> 0.1$. The opposite behaviour of dip and kink around $|\alpha_{21}|=0.04$ for both experiments is due to the presence of the additional NU phase $\phi_{21}$. As soon as we fix  $\phi_{21}=0$, the opposite nature of the experiments vanishes as shown by the dashed lines.   This indicates the NU phase $\phi_{21}$ exhibits extra degeneracy with standard oscillation parameters, thereby making the measurement of CPV sensitivity more challenging.


\subsection{Effect of NU parameters on the measurement of $\theta_{23}$ and $\Delta m^2_{31}$ }

\begin{figure}[h!]
\begin{center}
    \includegraphics[width=70mm, height=60mm]{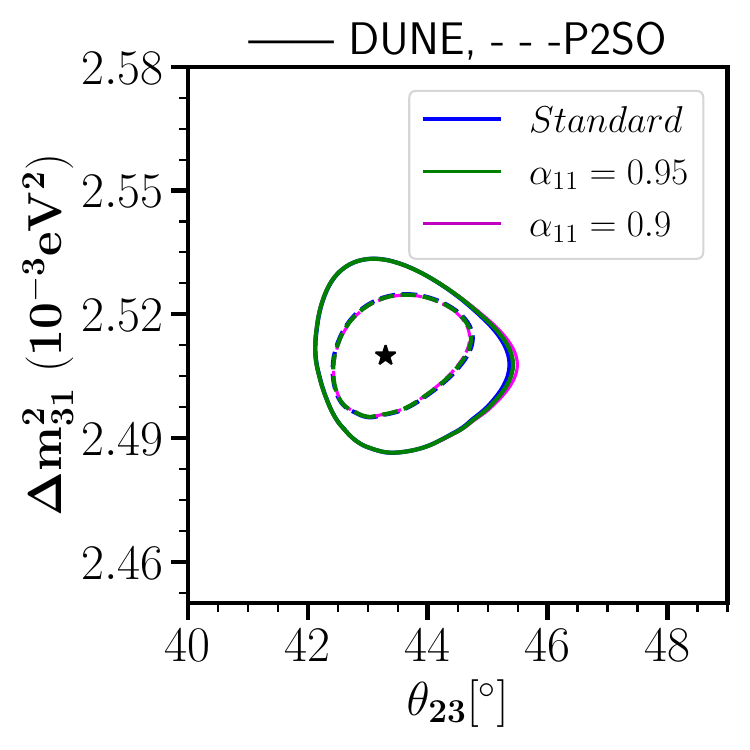}
\includegraphics[width=70mm, height=60mm]{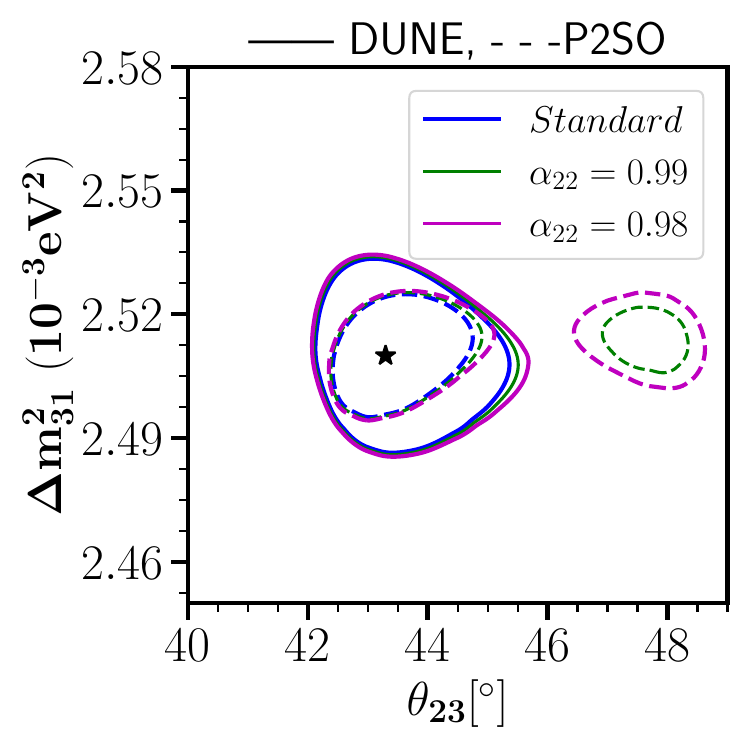}  \\

    \includegraphics[width=70mm, height=60mm]{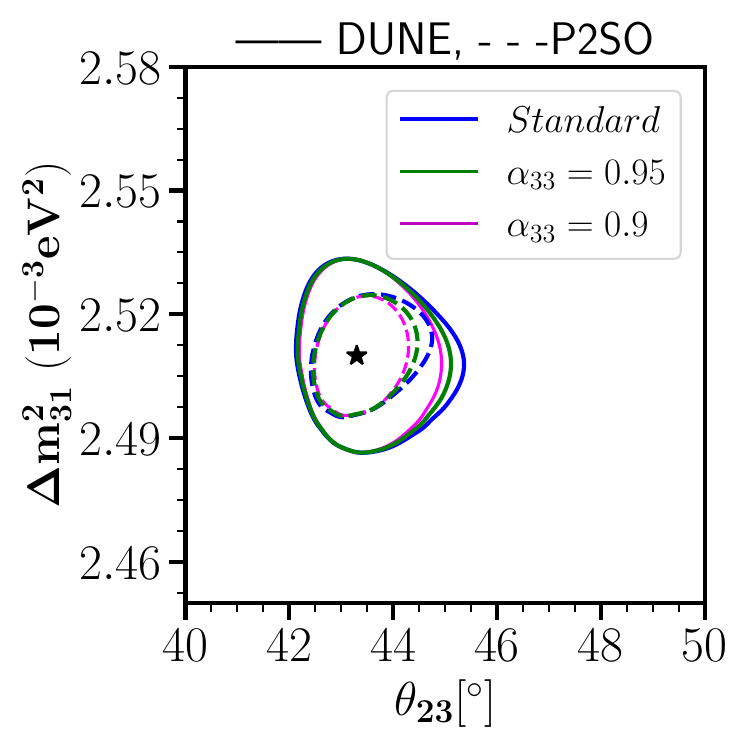}
    \includegraphics[width=70mm, height=60mm]{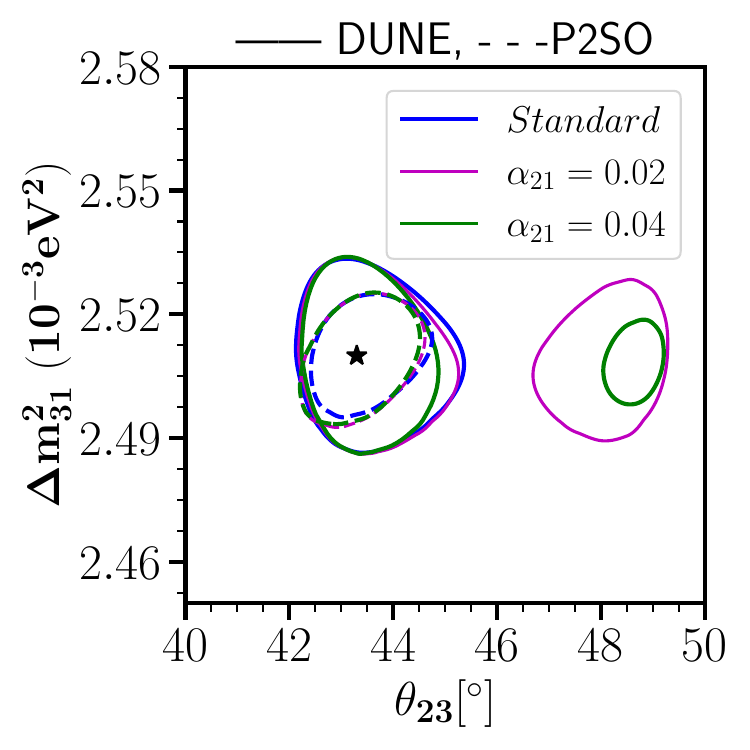}  

    \caption{Allowed regions in the $\theta_{23}$–$\Delta m^2_{31}$ plane at $3\sigma$ C.L. Two sets of $\alpha_{ij}$ are fixed in both the true and test spectra, as indicated in the legend, and compared with the standard scenario.}
    \label{th23ldm}
    \end{center}
\end{figure}

\begin{figure}
    \centering

    \includegraphics[width=80mm, height=65mm]{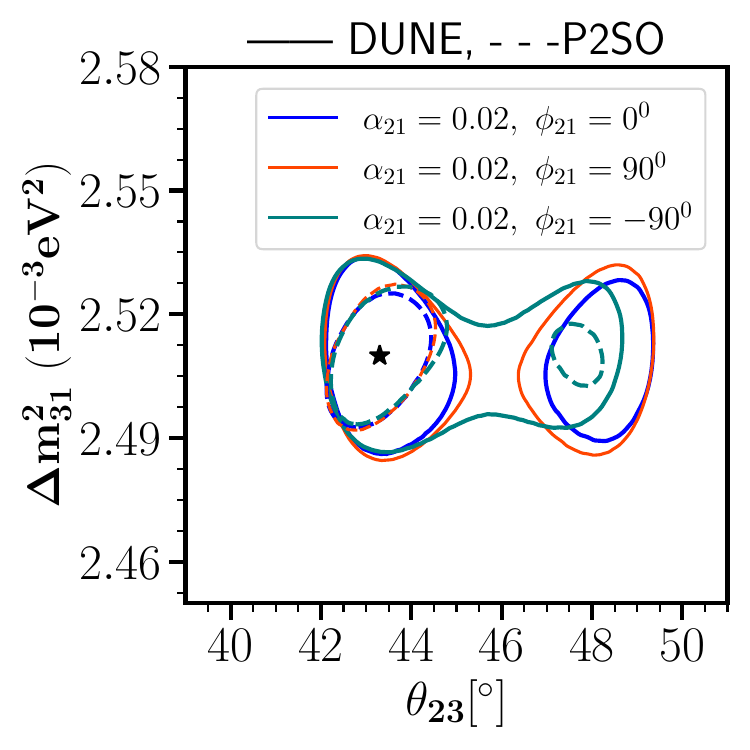}

    \caption{Allowed regions in the $\theta_{23}$–$\Delta m^2_{31}$ plane at $3\sigma$ C.L.   for fixed $\alpha_{21}$ values with three different true $\phi_{21}$ values.}
    \label{precision_alp21}
\end{figure}

In this subsection, we study the precision measurement of $\Delta m_{31}^2$ and $\theta_{23}$ assuming NU exits in Nature. Figure~\ref{th23ldm} shows the allowed regions of $\Delta m_{31}^2$ and $\theta_{23}$ at the $3\sigma$ confidence level for different NU parameters: $\alpha_{11}$ (upper-left), $\alpha_{22}$ (upper-right), $\alpha_{33}$ (lower-left), and $\alpha_{21}$ (lower-right). To generate this figure, we use the same minimization procedure described in Section \ref{res}. In the lower-right panel, the true value of the complex phase $\phi_{21}$ is taken as $ 0^\circ$. In each panel, the black star indicates the true values of $\Delta m_{31}^2$ and $\theta_{23}$ taken from Table~\ref{spara}. The blue contour represents the $3\sigma$ allowed region in the standard case i.e., $\alpha_{ii}=1$ and $\alpha_{ij}=0$. The green and magenta contours represent two distinct benchmark values for both the diagonal and off-diagonal NU parameters. These benchmark points are chosen such that they lie within their current allowed values and therefore vary from parameter to parameter. The dashed contours show the results for the P2SO experiment, while the solid contours represent the DUNE experiment.

We first discuss the effect of diagonal NU parameters. From the upper-left panel, we see that the presence of $\alpha_{11}$ does not significantly change the allowed standard parameter region depicted by the blue contour. However, in the presence of $\alpha_{22}$, shown in the upper-right panel, the size of the contour increases noticeably. In the DUNE setup, the green contour is larger than the blue, and the magenta contour expands further, indicating a progressive loss of sensitivity with increasing NU parameter. The precision on $\theta_{23}$ is particularly affected, in line with the decreasing octant sensitivity seen in the upper-right panel of Fig.~\ref{octsen}. On the contrary, P2SO exhibits degenerate regions in $\theta_{23}$, indicating reduced sensitivity for that NU benchmark points. This behavior, consistent with the dip present in upper right panel of Fig.~\ref{octsen}. This difference comes from the strong matter effects in P2SO compared to DUNE. Next, we consider the lower-left panel, which shows the effect of the $\alpha_{33}$ parameter. In this case, the precision of $\theta_{23}$ improves in the presence of $\alpha_{33}$. This improvement can be understood from the lower-left panel of Fig. \ref{octsen}, where the octant sensitivity increases when $\alpha_{33}$ is included.

For the off-diagonal NU parameter $\alpha_{21}$, in the P2SO setup, no degenerate regions in $\theta_{23}$ are observed. The corresponding contours shrink continuously as the NU strength increases. This trend is consistent with the smooth rise in octant sensitivity except a small dip as seen in the lower-right panel of Fig.~\ref{octsen}. In contrast, the DUNE experiment exhibits a pronounced dip in octant sensitivity for specific values of $\alpha_{21}$. As the benchmark points are chosen from these dip regions in Fig.~\ref{octsen}, the resulting contours show degenerate regions in $\theta_{23}$. This leads to an apparent loss of precision for those particular NU benchmarks, despite increasing $\alpha_{21}$. 
The emergence of these degenerate regions in DUNE, as opposed to their absence in P2SO, highlights the  roles played by $\phi_{21}$ in the two experimental configurations.

To further examine the role of complex phase ($\phi_{21}$), we show Fig.~\ref{precision_alp21} where we fix the magnitude of the NU parameter $\alpha_{21}$ and vary its phase, considering $\phi_{21}$=$ 0$, $90^\circ$ and $-90^\circ$. The resulting contours demonstrate a clear phase dependence in the sensitivity to $\theta_{23}$ for both experimental setups. For DUNE, $\phi_{21}=\pm90^\circ$ lead to an expansion of the contours compared to the CP-conserving case $\phi_{21}= 0$. P2SO experiment exhibits a milder dependence on $\phi_{21}$ in this corelation plot. The contours remain comparatively compact even in the presence of maximal CP-violating phases. However, there appears a degenerate region for $\phi_{21}=-90^\circ$. This figure indicate a degradation in precision driven by interference between standard oscillation terms and the extra phase associated with complex off-diagonal NU parameter.

\subsection{Allowed parameter space between $\deltaCP$ - $\phi_{21}$}

\begin{figure}[h!]
\begin{center}
    \includegraphics[width=75mm, height=62mm]{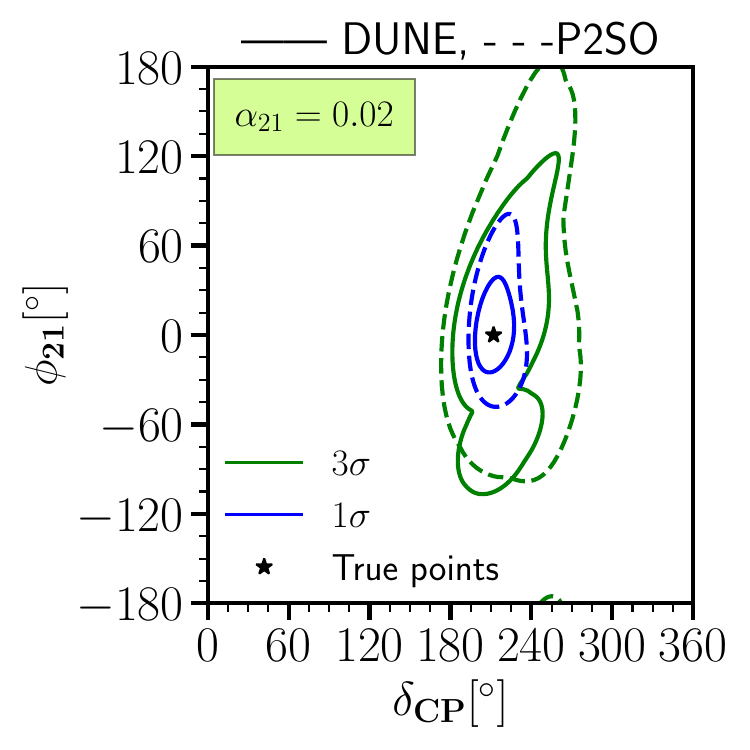}
    \includegraphics[width=75mm, height=62mm]{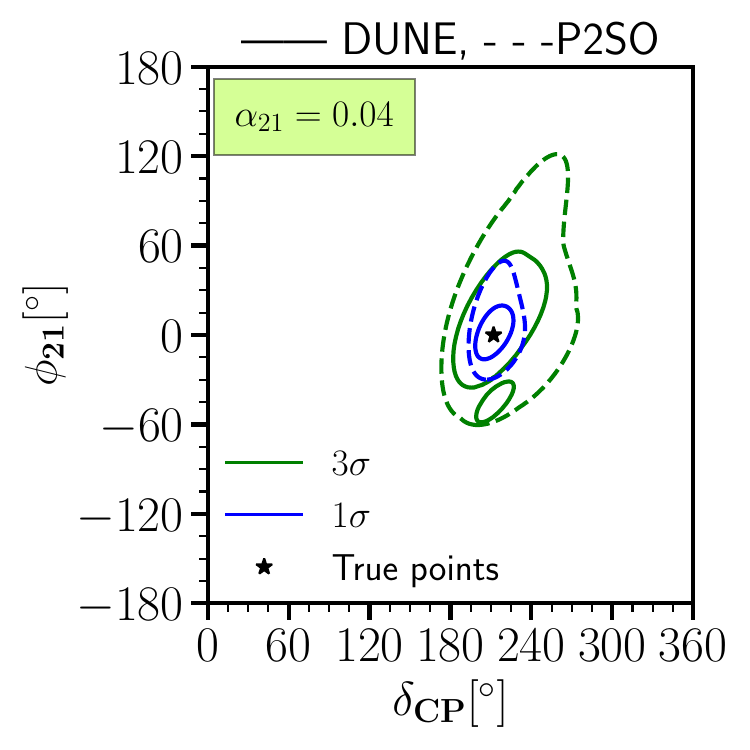}  

    \caption{Allowed parameter space between $\deltaCP$ (test) - $\phi_{21}$ (test) at $1\sigma$ and $3\sigma$ C.L. We give two panels for two benchmark values for the NU parameter $\alpha_{21}$.}
    \label{dcpphi}
    \end{center}
\end{figure}

As mentioned earlier, the complex phase associated with the off-diagonal NU parameter can affect the leptonic CP phase $\delta_{\rm CP}$. Therefore, it is important to study the correlation between $\delta_{\rm CP}$ and the NU complex phase $\phi_{21}$. In this subsection, we present the allowed parameter space in the $\delta_{\rm CP}$–$\phi_{21}$ plane for two fixed values of the off-diagonal NU parameter $\alpha_{21}$ assuming the true values for $\delta_{\rm CP}$ and $\phi_{21}$ to be $212^\circ$ and $0$ respectively. Figure~\ref{dcpphi} shows how $\phi_{21}$ depends on $\delta_{\rm CP}$ for $\alpha_{21} = 0.02$, and $0.04$.  To generate this figure, the true value of $\phi_{21}$ is fixed at $0$, while in the test spectrum it is varied over the range $-180^\circ$ to $180^\circ$. In each panel, the blue and green contours represent the $1\sigma$ and $3\sigma$ confidence level allowed regions, respectively. The solid contours correspond to the DUNE experiment, while the dashed contours represent the P2SO experiment. From the figure we notice that the precision of $\phi_{21}$ is generally weaker than the precision $\delta_{\rm CP}$ for the chosen values. As we increase the value of  $\alpha_{21}$, the precision of $\phi_{21}$ improves. Between the two experiments, DUNE provides stronger constraints than P2SO.


\section{Summary and conclusion}
\label{conclusion}

In this work, we perform a detailed comparative study of NU neutrino mixing in the context of the upcoming long-baseline experiments P2SO and DUNE. The technical specifications adopted in our simulations are summarized in section~\ref{sim}. We derive bounds on the NU parameters under two scenarios: (i) varying one parameter at a time (one dof), and (ii) allowing all relevant NU parameters to vary simultaneously (six dof).
We find that the two experiments provide comparable sensitivities overall, with complementary strengths. DUNE yields stronger constraints on $\alpha_{11}$ and $\alpha_{21}$, whereas P2SO shows enhanced sensitivity to $\alpha_{22}$ and $\alpha_{33}$. Our analysis shows that DUNE will be able to improve the current limit on $\alpha_{11}$ whereas P2SO will be able to improve the current limit on $\alpha_{33}$. The bounds on $\alpha_{33}$ strongly depends upon the parameter $\theta_{23}$ and the bound on the off-diagonal parameter $\alpha_{21}$ depends significantly on the phase $\phi_{21}$. Further, we have also shown that for marginalizing over $\Delta m^2_{31}$ within its present global $3 \sigma$ range is sufficient to constrain the NU parameters precisely.

We further investigate the impact of NU parameters on the sensitivities to the neutrino mass hierarchy, the octant of $\theta_{23}$, and leptonic CP violation. In the presence of $\alpha_{11}$, the mass hierarchy sensitivity is notably reduced relative to the standard three-flavor expectation, although it remains well above $5\sigma$. In contrast, $\alpha_{22}$ enhances the hierarchy sensitivity, primarily through modifications in the disappearance channel. The dependence on $\alpha_{33}$ and $\alpha_{21}$ is comparatively mild. The octant sensitivity is also significantly affected by NU. The parameter $\alpha_{11}$ does not affect the sensitivity much, however, $\alpha_{22}$ reduce the sensitivity, whereas $\alpha_{33}$ leads to an enhancement. We also demonstrated that the dip in the octant sensitivity around $\alpha_{22} = 0.97$ is due the matter effect. The additional phase $\phi_{21}$ plays a crucial role in the context of octant sensitivity in presence of $\alpha_{21}$. For small values of $|\alpha_{21}|$, the octant sensitivity deteriorates relative to the standard case due the degeneracy associated with $\phi_{21}$. However, for larger magnitudes of $|\alpha_{21}|$, the sensitivity increases from standard prediction. The CP violation sensitivity decreases in presence of $\alpha_{11}$ and  $\alpha_{22}$ whereas it increases in presence of $\alpha_{33}$. The CPV sensitivity in presence of $\alpha_{21}$ is a bit non-trivial and it depends upon the phase $\phi_{21}$.

Finally, we present precision contours in the $\theta_{23}$–$\Delta m^2_{31}$ plane. We observe that $\alpha_{22}$ and $\alpha_{21}$ induce the largest distortions of the allowed regions giving rise to disjoint degenerate regions for $\alpha_{22}$ in P2SO and $\alpha_{21}$ in DUNE due to octant degeneracy. The effects of $\alpha_{11}$ and $\alpha_{33}$ are comparatively moderate. For $\alpha_{21}$, we tried to see the effect of the phase $\phi_{21}$ and showed that maximal values of $\phi_{21}$ tends to deteriorate the sensitivity as compared to $\phi_{21} = 0$. Finally, we also studied the precision of the two phases $\delta_{\rm CP}$ and $\phi_{21}$, in presence of $\alpha_{21}$. We notice that the precision of $\phi_{21}$ is generally weaker than the precision $\delta_{\rm CP}$ for our chosen values. As we increase the value of  $\alpha_{21}$, the precision of $\phi_{21}$ improves.

Overall, both DUNE and P2SO can place stringent constraints on NU parameters, and NU effects can non-trivially impact the determination of the currently unknown oscillation parameters. The complementarity of the two experiments is therefore essential for a robust resolution of these unknowns in the presence of non-unitarity.

\acknowledgments
We thank Papia Panda for many useful discussions and carefully reading the manuscript.
SKP thanks the University Grants Commission for the NFOBC fellowship.
SR is supported by the NPDF grant (PDF/2023/001262) from ANRF,
Government of India. The work of MG has been in part funded by Ministry of Science and Education of Republic of Croatia grant No. PK.1.1.10.0002, Swiss National Science Foundation (SNSF) and Croatian Science Foundation (HRZZ) under grant MAPS IZ11Z0$\_$230193 and European Union under the NextGenerationEU Programme. Views and opinions expressed are however those of
the author(s) only and do not necessarily reflect those of the European Union. Neither the
European Union nor the granting authority can be held responsible for them. We gratefully
acknowledge the use of CMSD HPC facility of University of Hyderabad to carry out the
computational works.


\bibliography{mybib}

@article{Fukugita:1986hr,
    author = "Fukugita, M. and Yanagida, T.",
    title = "{Baryogenesis Without Grand Unification}",
    reportNumber = "RIFP-641",
    doi = "10.1016/0370-2693(86)91126-3",
    journal = "Phys. Lett. B",
    volume = "174",
    pages = "45--47",
    year = "1986"
}

@article{Maki:1962mu,
    author = "Maki, Ziro and Nakagawa, Masami and Sakata, Shoichi",
    title = "{Remarks on the unified model of elementary particles}",
    doi = "10.1143/PTP.28.870",
    journal = "Prog. Theor. Phys.",
    volume = "28",
    pages = "870--880",
    year = "1962"
}

@article{DUNE:2020jqi,
    author = "Abi, B. and others",
    collaboration = "DUNE",
    title = "{Long-baseline neutrino oscillation physics potential of the DUNE experiment}",
    eprint = "2006.16043",
    archivePrefix = "arXiv",
    primaryClass = "hep-ex",
    reportNumber = "FERMILAB-PUB-20-251-E-LBNF-ND-PIP2-SCD, PUB-20-251-E-LBNF-ND-PIP2-SCD",
    doi = "10.1140/epjc/s10052-020-08456-z",
    journal = "Eur. Phys. J. C",
    volume = "80",
    number = "10",
    pages = "978",
    year = "2020"
}

@article{Fogli:2002pt,
    author = "Fogli, G. L. and Lisi, E. and Marrone, A. and Montanino, D. and Palazzo, A.",
    title = "{Getting the most from the statistical analysis of solar neutrino oscillations}",
    eprint = "hep-ph/0206162",
    archivePrefix = "arXiv",
    doi = "10.1103/PhysRevD.66.053010",
    journal = "Phys. Rev. D",
    volume = "66",
    pages = "053010",
    year = "2002"
}

@article{Huber:2002mx,
    author = "Huber, Patrick and Lindner, Manfred and Winter, Walter",
    title = "{Superbeams versus neutrino factories}",
    eprint = "hep-ph/0204352",
    archivePrefix = "arXiv",
    reportNumber = "TUM-HEP-462-02, MPI-PHT-02-15",
    doi = "10.1016/S0550-3213(02)00825-8",
    journal = "Nucl. Phys. B",
    volume = "645",
    pages = "3--48",
    year ="2002"
}

@article{Joshipura:2001ui,
    author = "Joshipura, Anjan S. and Paschos, Emmanuel A. and Rodejohann, Werner",
    title = "{A Simple connection between neutrino oscillation and leptogenesis}",
    eprint = "hep-ph/0105175",
    archivePrefix = "arXiv",
    reportNumber = "DO-TH-01-06",
    doi = "10.1088/1126-6708/2001/08/029",
    journal = "JHEP",
    volume = "08",
    pages = "029",
    year = "2001"
}

@article{NOvA:2021nfi,
    author = "Acero, M. A. and others",
    collaboration = "NOvA",
    title = "{Improved measurement of neutrino oscillation parameters by the NOvA experiment}",
    eprint = "2108.08219",
    archivePrefix = "arXiv",
    primaryClass = "hep-ex",
    reportNumber = "FERMILAB-PUB-21-373-ND",
    doi = "10.1103/PhysRevD.106.032004",
    journal = "Phys. Rev. D",
    volume = "106",
    number = "3",
    pages = "032004",
    year = "2022"
}

@article{T2K:2025yoy,
    author = "Abe, K. and others",
    collaboration = "T2K",
    title = "{Results from the T2K experiment on neutrino mixing including a new far detector $\mu$-like sample}",
    eprint = "2506.05889",
    archivePrefix = "arXiv",
    primaryClass = "hep-ex",
    month = "6",
    year = "2025"
}

@article{IceCubeCollaboration:2024ssx,
    author = "Abbasi, R. and others",
    collaboration = "(IceCube Collaboration){\ensuremath{\parallel}}, IceCube",
    title = "{Measurement of Atmospheric Neutrino Oscillation Parameters Using Convolutional Neural Networks with 9.3 Years of Data in IceCube DeepCore}",
    eprint = "2405.02163",
    archivePrefix = "arXiv",
    primaryClass = "hep-ex",
    doi = "10.1103/PhysRevLett.134.091801",
    journal = "Phys. Rev. Lett.",
    volume = "134",
    number = "9",
    pages = "091801",
    year = "2025"
}

@article{Endoh:2002wm,
    author = "Endoh, T. and Kaneko, S. and Kang, S. K. and Morozumi, T. and Tanimoto, M.",
    title = "{CP violation in neutrino oscillation and leptogenesis}",
    eprint = "hep-ph/0209020",
    archivePrefix = "arXiv",
    reportNumber = "HUPD-0205",
    doi = "10.1103/PhysRevLett.89.231601",
    journal = "Phys. Rev. Lett.",
    volume = "89",
    pages = "231601",
    year = "2002"
}

@article{Buchmuller:2005eh,
    author = "Buchmuller, W. and Peccei, R. D. and Yanagida, T.",
    title = "{Leptogenesis as the origin of matter}",
    eprint = "hep-ph/0502169",
    archivePrefix = "arXiv",
    reportNumber = "DESY-05-031",
    doi = "10.1146/annurev.nucl.55.090704.151558",
    journal = "Ann. Rev. Nucl. Part. Sci.",
    volume = "55",
    pages = "311--355",
    year = "2005"
}

@article{KATRIN:2024cdt,
    author = "Aker, Max and others",
    collaboration = "KATRIN",
    title = "{Direct neutrino-mass measurement based on 259 days of KATRIN data}",
    eprint = "2406.13516",
    archivePrefix = "arXiv",
    primaryClass = "nucl-ex",
    doi = "10.1126/science.adq9592",
    journal = "Science",
    volume = "388",
    number = "6743",
    pages = "adq9592",
    year = "2025"
}

@article{DESI:2024hhd,
    author = "Adame, A. G. and others",
    collaboration = "DESI",
    title = "{DESI 2024 VII: cosmological constraints from the full-shape modeling of clustering measurements}",
    eprint = "2411.12022",
    archivePrefix = "arXiv",
    primaryClass = "astro-ph.CO",
    reportNumber = "FERMILAB-PUB-24-0854-PPD",
    doi = "10.1088/1475-7516/2025/07/028",
    journal = "JCAP",
    volume = "07",
    pages = "028",
    year = "2025"
}

@article{Shao:2024mag,
    author = "Shao, Helen and Givans, Jahmour J. and Dunkley, Jo and Madhavacheril, Mathew and Qu, Frank J. and Farren, Gerrit and Sherwin, Blake",
    title = "{Cosmological limits on the neutrino mass sum for beyond-{\ensuremath{\Lambda}}CDM models}",
    eprint = "2409.02295",
    archivePrefix = "arXiv",
    primaryClass = "astro-ph.CO",
    doi = "10.1103/PhysRevD.111.083535",
    journal = "Phys. Rev. D",
    volume = "111",
    number = "8",
    pages = "083535",
    year = "2025"
}

@article{DESI:2025gwf,
    author = "Garcia-Quintero, C. and others",
    collaboration = "DESI",
    title = "{Cosmological implications of DESI DR2 BAO measurements in light of the latest ACT DR6 CMB data}",
    eprint = "2504.18464",
    archivePrefix = "arXiv",
    primaryClass = "astro-ph.CO",
    reportNumber = "FERMILAB-PUB-25-0281-PPD",
    month = "4",
    year = "2025"
}

@article{McDonald:2016ixn,
    author = "McDonald, Arthur B.",
    title = "{Nobel Lecture: The Sudbury Neutrino Observatory: Observation of flavor change for solar neutrinos}",
    doi = "10.1103/RevModPhys.88.030502",
    journal = "Rev. Mod. Phys.",
    volume = "88",
    number = "3",
    pages = "030502",
    year = "2016"
}

@article{Kajita:2016cak,
    author = "Kajita, Takaaki",
    title = "{Nobel Lecture: Discovery of atmospheric neutrino oscillations}",
    doi = "10.1103/RevModPhys.88.030501",
    journal = "Rev. Mod. Phys.",
    volume = "88",
    number = "3",
    pages = "030501",
    year = "2016"
}

@article{Abazajian:2012ys,
    author = "Abazajian, K. N. and others",
    title = "{Light Sterile Neutrinos: A White Paper}",
    eprint = "1204.5379",
    archivePrefix = "arXiv",
    primaryClass = "hep-ph",
    reportNumber = "FERMILAB-PUB-12-881-PPD",
    month = "4",
    year = "2012"
}

@article{Gronau:1984ct,
    author = "Gronau, Michael and Leung, Chung Ngoc and Rosner, Jonathan L.",
    title = "{Extending Limits on Neutral Heavy Leptons}",
    reportNumber = "EFI-83-63-CHICAGO, FERMILAB-PUB-84-024-T",
    doi = "10.1103/PhysRevD.29.2539",
    journal = "Phys. Rev. D",
    volume = "29",
    pages = "2539",
    year = "1984"
}

@article{Nardi:1994iv,
    author = "Nardi, Enrico and Roulet, Esteban and Tommasini, Daniele",
    title = "{Limits on neutrino mixing with new heavy particles}",
    eprint = "hep-ph/9402224",
    archivePrefix = "arXiv",
    reportNumber = "FTUV-93-47, UM-TH-93-28, CERN-TH-7150-94",
    doi = "10.1016/0370-2693(94)90736-6",
    journal = "Phys. Lett. B",
    volume = "327",
    pages = "319--326",
    year = "1994"
}

@article{Atre:2009rg,
    author = "Atre, Anupama and Han, Tao and Pascoli, Silvia and Zhang, Bin",
    title = "{The Search for Heavy Majorana Neutrinos}",
    eprint = "0901.3589",
    archivePrefix = "arXiv",
    primaryClass = "hep-ph",
    reportNumber = "FERMILAB-PUB-08-086-T, NSF-KITP-08-54, MADPH-06-1466, DCPT-07-198, IPPP-07-99",
    doi = "10.1088/1126-6708/2009/05/030",
    journal = "JHEP",
    volume = "05",
    pages = "030",
    year = "2009"
}

@article{Mohapatra:1979ia,
    author = "Mohapatra, Rabindra N. and Senjanovic, Goran",
    title = "{Neutrino Mass and Spontaneous Parity Nonconservation}",
    reportNumber = "MDDP-TR-80-060, MDDP-PP-80-105, CCNY-HEP-79-10",
    doi = "10.1103/PhysRevLett.44.912",
    journal = "Phys. Rev. Lett.",
    volume = "44",
    pages = "912",
    year = "1980"
}

@article{Gell-Mann:1979vob,
    author = "Gell-Mann, Murray and Ramond, Pierre and Slansky, Richard",
    title = "{Complex Spinors and Unified Theories}",
    eprint = "1306.4669",
    archivePrefix = "arXiv",
    primaryClass = "hep-th",
    reportNumber = "PRINT-80-0576",
    journal = "Conf. Proc. C",
    volume = "790927",
    pages = "315--321",
    year = "1979"
}

@article{Schechter:1980gr,
    author = "Schechter, J. and Valle, J. W. F.",
    title = "{Neutrino Masses in SU(2) x U(1) Theories}",
    reportNumber = "SU-4217-167, COO-3533-167",
    doi = "10.1103/PhysRevD.22.2227",
    journal = "Phys. Rev. D",
    volume = "22",
    pages = "2227",
    year = "1980"
}

@book{Valle:2015pba,
    author = "Valle, Jose W. F. and Romao, Jorge C.",
    title = "{Neutrinos in high energy and astroparticle physics}",
    isbn = "978-3-527-41197-9, 978-3-527-67102-1",
    publisher = "Wiley-VCH",
    address = "Weinheim",
    series = "Physics textbook",
    year = "2015"
}

@article{LSND:2001aii,
    author = "Aguilar, A. and others",
    collaboration = "LSND",
    title = "{Evidence for neutrino oscillations from the observation of $\bar{\nu}_e$ appearance in a $\bar{\nu}_\mu$
 beam}",
    eprint = "hep-ex/0104049",
    archivePrefix = "arXiv",
    doi = "10.1103/PhysRevD.64.112007",
    journal = "Phys. Rev. D",
    volume = "64",
    pages = "112007",
    year = "2001"
}

@article{MiniBooNE:2018esg,
    author = "Aguilar-Arevalo, A. A. and others",
    collaboration = "MiniBooNE",
    title = "{Significant Excess of ElectronLike Events in the MiniBooNE Short-Baseline Neutrino Experiment}",
    eprint = "1805.12028",
    archivePrefix = "arXiv",
    primaryClass = "hep-ex",
    reportNumber = "LA-UR-18-24586, FERMILAB-PUB-18-219-AD-PPD-ND",
    doi = "10.1103/PhysRevLett.121.221801",
    journal = "Phys. Rev. Lett.",
    volume = "121",
    number = "22",
    pages = "221801",
    year = "2018"
}

@article{Mention:2011rk,
    author = "Mention, G. and Fechner, M. and Lasserre, Th. and Mueller, Th. A. and Lhuillier, D. and Cribier, M. and Letourneau, A.",
    title = "{The Reactor Antineutrino Anomaly}",
    eprint = "1101.2755",
    archivePrefix = "arXiv",
    primaryClass = "hep-ex",
    doi = "10.1103/PhysRevD.83.073006",
    journal = "Phys. Rev. D",
    volume = "83",
    pages = "073006",
    year = "2011"
}

@article{Huber:2011wv,
    author = "Huber, Patrick",
    title = "{On the determination of anti-neutrino spectra from nuclear reactors}",
    eprint = "1106.0687",
    archivePrefix = "arXiv",
    primaryClass = "hep-ph",
    doi = "10.1103/PhysRevC.85.029901",
    journal = "Phys. Rev. C",
    volume = "84",
    pages = "024617",
    year = "2011",
    note = "[Erratum: Phys.Rev.C 85, 029901 (2012)]"
}

@article{Fernandez-Martinez:2016lgt,
    author = "Fernandez-Martinez, Enrique and Hernandez-Garcia, Josu and Lopez-Pavon, Jacobo",
    title = "{Global constraints on heavy neutrino mixing}",
    eprint = "1605.08774",
    archivePrefix = "arXiv",
    primaryClass = "hep-ph",
    doi = "10.1007/JHEP08(2016)033",
    journal = "JHEP",
    volume = "08",
    pages = "033",
    year = "2016"
}

@article{Blennow:2025qgd,
    author = "Blennow, Mattias and Coloma, Pilar and Fern{\'a}ndez-Mart{\'\i}nez, Enrique and Hern{\'a}ndez-Garc{\'\i}a, Josu and L{\'o}pez-Pav{\'o}n, Jacobo and Marcano, Xabier and Naredo-Tuero, Daniel and Urrea, Salvador",
    title = "{Misconceptions in neutrino oscillations in presence of non-unitary mixing}",
    eprint = "2502.19480",
    archivePrefix = "arXiv",
    primaryClass = "hep-ph",
    reportNumber = "IFT-UAM/CSIC-25-20, FTUV-25-0225.3158",
    doi = "10.1016/j.nuclphysb.2025.116944",
    journal = "Nucl. Phys. B",
    volume = "1017",
    pages = "116944",
    year = "2025"
}

@article{Blennow:2023mqx,
    author = "Blennow, Mattias and Fern\'andez-Mart\'\i{}nez, Enrique and Hern\'andez-Garc\'\i{}a, Josu and L\'opez-Pav\'on, Jacobo and Marcano, Xabier and Naredo-Tuero, Daniel",
    title = "{Bounds on lepton non-unitarity and heavy neutrino mixing}",
    eprint = "2306.01040",
    archivePrefix = "arXiv",
    primaryClass = "hep-ph",
    reportNumber = "IFT-UAM/CSIC-23-60, FTUV-23-0531.7594, IFIC/23-19",
    doi = "10.1007/JHEP08(2023)030",
    journal = "JHEP",
    volume = "08",
    pages = "030",
    year = "2023"
}

@article{Forero:2021azc,
    author = "Forero, D. V. and Giunti, C. and Ternes, C. A. and Tortola, M.",
    title = "{Nonunitary neutrino mixing in short and long-baseline experiments}",
    eprint = "2103.01998",
    archivePrefix = "arXiv",
    primaryClass = "hep-ph",
    doi = "10.1103/PhysRevD.104.075030",
    journal = "Phys. Rev. D",
    volume = "104",
    number = "7",
    pages = "075030",
    year = "2021"
}

@article{Denton:2021mso,
    author = "Denton, Peter B. and Gehrlein, Julia",
    title = "{New oscillation and scattering constraints on the tau row matrix elements without assuming unitarity}",
    eprint = "2109.14575",
    archivePrefix = "arXiv",
    primaryClass = "hep-ph",
    doi = "10.1007/JHEP06(2022)135",
    journal = "JHEP",
    volume = "06",
    pages = "135",
    year = "2022"
}

@article{Dutta:2019hmb,
    author = "Dutta, Debajyoti and Roy, Samiran",
    title = "{Non-Unitarity at DUNE and T2HK with Charged and Neutral Current Measurements}",
    eprint = "1901.11298",
    archivePrefix = "arXiv",
    primaryClass = "hep-ph",
    doi = "10.1088/1361-6471/abdc03",
    journal = "J. Phys. G",
    volume = "48",
    number = "4",
    pages = "045004",
    year = "2021"
}

@article{Ge:2016xya,
    author = "Ge, Shao-Feng and Pasquini, Pedro and Tortola, M. and Valle, J. W. F.",
    title = "{Measuring the leptonic CP phase in neutrino oscillations with nonunitary mixing}",
    eprint = "1605.01670",
    archivePrefix = "arXiv",
    primaryClass = "hep-ph",
    reportNumber = "IFIC-16-26",
    doi = "10.1103/PhysRevD.95.033005",
    journal = "Phys. Rev. D",
    volume = "95",
    number = "3",
    pages = "033005",
    year = "2017"
}

@article{Meloni:2009cg,
    author = "Meloni, Davide and Ohlsson, Tommy and Winter, Walter and Zhang, He",
    title = "{Non-standard interactions versus non-unitary lepton flavor mixing at a neutrino factory}",
    eprint = "0912.2735",
    archivePrefix = "arXiv",
    primaryClass = "hep-ph",
    reportNumber = "NORDITA-2009-79, IDS-NF-014",
    doi = "10.1007/JHEP04(2010)041",
    journal = "JHEP",
    volume = "04",
    pages = "041",
    year = "2010"
}

@article{Blennow:2016jkn,
    author = "Blennow, Mattias and Coloma, Pilar and Fernandez-Martinez, Enrique and Hernandez-Garcia, Josu and Lopez-Pavon, Jacobo",
    title = "{Non-Unitarity, sterile neutrinos, and Non-Standard neutrino Interactions}",
    eprint = "1609.08637",
    archivePrefix = "arXiv",
    primaryClass = "hep-ph",
    reportNumber = "IFT-UAM-CSIC-16-090, FTUAM-16-35, FERMILAB-PUB-16-400-T",
    doi = "10.1007/JHEP04(2017)153",
    journal = "JHEP",
    volume = "04",
    pages = "153",
    year = "2017"
}

@article{Miranda:2018yym,
    author = "Miranda, O. G. and Pasquini, Pedro and T\'ortola, M. and Valle, J. W. F.",
    title = "{Exploring the Potential of Short-Baseline Physics at Fermilab}",
    eprint = "1802.02133",
    archivePrefix = "arXiv",
    primaryClass = "hep-ph",
    reportNumber = "IFIC-18-XXX",
    doi = "10.1103/PhysRevD.97.095026",
    journal = "Phys. Rev. D",
    volume = "97",
    number = "9",
    pages = "095026",
    year = "2018"
}

@article{Miranda:2020syh,
    author = "Miranda, O. G. and Papoulias, D. K. and Sanders, O. and T\'ortola, M. and Valle, J. W. F.",
    title = "{Future CEvNS experiments as probes of lepton unitarity and light-sterile neutrinos}",
    eprint = "2008.02759",
    archivePrefix = "arXiv",
    primaryClass = "hep-ph",
    doi = "10.1103/PhysRevD.102.113014",
    journal = "Phys. Rev. D",
    volume = "102",
    pages = "113014",
    year = "2020"
}

@article{C:2017scx,
    author = "C, Soumya and Mohanta, Rukmani",
    title = "{Non-unitary lepton mixing in an inverse seesaw and its impact on the physics potential of long-baseline experiments}",
    eprint = "1708.05372",
    archivePrefix = "arXiv",
    primaryClass = "hep-ph",
    month = "8",
    year = "2017"
}

@article{Coloma:2021uhq,
    author = "Coloma, Pilar and L\'opez-Pav\'on, Jacobo and Rosauro-Alcaraz, Salvador and Urrea, Salvador",
    title = "{New physics from oscillations at the DUNE near detector, and the role of systematic uncertainties}",
    eprint = "2105.11466",
    archivePrefix = "arXiv",
    primaryClass = "hep-ph",
    reportNumber = "FTUV-21-0505.4636, IFT-UAM/CSIC-21-51, IFIC/21-14",
    doi = "10.1007/JHEP08(2021)065",
    journal = "JHEP",
    volume = "08",
    pages = "065",
    year = "2021"
}

@article{Agarwalla:2021owd,
    author = "Agarwalla, Sanjib Kumar and Das, Sudipta and Giarnetti, Alessio and Meloni, Davide",
    title = "{Model-independent constraints on non-unitary neutrino mixing from high-precision long-baseline experiments}",
    eprint = "2111.00329",
    archivePrefix = "arXiv",
    primaryClass = "hep-ph",
    reportNumber = "IP/BBSR/2021-10",
    doi = "10.1007/JHEP07(2022)121",
    journal = "JHEP",
    volume = "07",
    pages = "121",
    year = "2022"
}

@article{Fernandez-Martinez:2007iaa,
    author = "Fernandez-Martinez, E. and Gavela, M. B. and Lopez-Pavon, J. and Yasuda, O.",
    title = "{CP-violation from non-unitary leptonic mixing}",
    eprint = "hep-ph/0703098",
    archivePrefix = "arXiv",
    reportNumber = "FTUAM-07-4, IFT-UAM-CSIC-07-10",
    doi = "10.1016/j.physletb.2007.03.069",
    journal = "Phys. Lett. B",
    volume = "649",
    pages = "427--435",
    year = "2007"
}

@article{Xing:2011ur,
    author = "Xing, Zhi-zhong",
    title = "{A full parametrization of the 6 X 6 flavor mixing matrix in the presence of three light or heavy sterile neutrinos}",
    eprint = "1110.0083",
    archivePrefix = "arXiv",
    primaryClass = "hep-ph",
    doi = "10.1103/PhysRevD.85.013008",
    journal = "Phys. Rev. D",
    volume = "85",
    pages = "013008",
    year = "2012"
}

@article{Escrihuela:2015wra,
    author = "Escrihuela, F. J. and Forero, D. V. and Miranda, O. G. and Tortola, M. and Valle, J. W. F.",
    title = "{On the description of nonunitary neutrino mixing}",
    eprint = "1503.08879",
    archivePrefix = "arXiv",
    primaryClass = "hep-ph",
    reportNumber = "IFIC-15-14",
    doi = "10.1103/PhysRevD.92.053009",
    journal = "Phys. Rev. D",
    volume = "92",
    number = "5",
    pages = "053009",
    year = "2015",
    note = "[Erratum: Phys.Rev.D 93, 119905 (2016)]"
}

@article{Esteban:2024eli,
    author = "Esteban, Ivan and Gonzalez-Garcia, M. C. and Maltoni, Michele and Martinez-Soler, Ivan and Pinheiro, Jo{\~a}o Paulo and Schwetz, Thomas",
    title = "{NuFit-6.0: updated global analysis of three-flavor neutrino oscillations}",
    eprint = "2410.05380",
    archivePrefix = "arXiv",
    primaryClass = "hep-ph",
    reportNumber = "IFT-UAM/CSIC-24-140, YITP-SB-2024-24, IPPP/24/64, IPPP/24/64, IFT-UAM/CSIC-24-140, YITP-SB-2024-24",
    doi = "10.1007/JHEP12(2024)216",
    journal = "JHEP",
    volume = "12",
    pages = "216",
    year = "2024"
}

@article{Akindinov:2019flp,
    author = "Akindinov, A. V. and others",
    title = "{Letter of Interest for a Neutrino Beam from Protvino to KM3NeT/ORCA}",
    eprint = "1902.06083",
    archivePrefix = "arXiv",
    primaryClass = "physics.ins-det",
    doi = "10.1140/epjc/s10052-019-7259-5",
    journal = "Eur. Phys. J. C",
    volume = "79",
    number = "9",
    pages = "758",
    year = "2019"
}

@article{Singha:2022btw,
    author = "Singha, Dinesh Kumar and Ghosh, Monojit and Majhi, Rudra and Mohanta, Rukmani",
    title = "{Study of light sterile neutrino at the long-baseline experiment options at KM3NeT}",
    eprint = "2211.01816",
    archivePrefix = "arXiv",
    primaryClass = "hep-ph",
    doi = "10.1103/PhysRevD.107.075039",
    journal = "Phys. Rev. D",
    volume = "107",
    number = "7",
    pages = "075039",
    year = "2023"
}

@article{Majhi:2022fed,
    author = "Majhi, Rudra and Singha, Dinesh Kumar and Ghosh, Monojit and Mohanta, Rukmani",
    title = "{Distinguishing nonstandard interaction and Lorentz invariance violation at the Protvino to super-ORCA experiment}",
    eprint = "2212.07244",
    archivePrefix = "arXiv",
    primaryClass = "hep-ph",
    doi = "10.1103/PhysRevD.107.075036",
    journal = "Phys. Rev. D",
    volume = "107",
    number = "7",
    pages = "075036",
    year = "2023"
}

@article{Singha:2023set,
    author = "Singha, Dinesh Kumar and Majhi, Rudra and Panda, Lipsarani and Ghosh, Monojit and Mohanta, Rukmani",
    title = "{Study of scalar nonstandard interaction at the Protvino to super-ORCA experiment}",
    eprint = "2308.10789",
    archivePrefix = "arXiv",
    primaryClass = "hep-ph",
    doi = "10.1103/PhysRevD.109.095038",
    journal = "Phys. Rev. D",
    volume = "109",
    number = "9",
    pages = "095038",
    year = "2024"
}

@article{DUNE:2021cuw,
    author = "Abi, B. and others",
    collaboration = "DUNE",
    title = "{Experiment Simulation Configurations Approximating DUNE TDR}",
    eprint = "2103.04797",
    archivePrefix = "arXiv",
    primaryClass = "hep-ex",
    reportNumber = "FERMILAB-FN-1125-ND",
    month = "3",
    year = "2021"
}

@article{Goswami:2008mi,
    author = "Goswami, Srubabati and Ota, Toshihiko",
    title = "{Testing non-unitarity of neutrino mixing matrices at neutrino factories}",
    eprint = "0802.1434",
    archivePrefix = "arXiv",
    primaryClass = "hep-ph",
    doi = "10.1103/PhysRevD.78.033012",
    journal = "Phys. Rev. D",
    volume = "78",
    pages = "033012",
    year = "2008"
}

@article{Chatterjee:2021xyu,
    author = "Chatterjee, Sabya Sachi and Miranda, O. G. and T{\'o}rtola, M. and Valle, J. W. F.",
    title = "{Nonunitarity of the lepton mixing matrix at the European Spallation Source}",
    eprint = "2111.08673",
    archivePrefix = "arXiv",
    primaryClass = "hep-ph",
    reportNumber = "t21/073",
    doi = "10.1103/PhysRevD.106.075016",
    journal = "Phys. Rev. D",
    volume = "106",
    number = "7",
    pages = "075016",
    year = "2022"
}

@article{Dutta:2016czj,
    author = "Dutta, Debajyoti and Ghoshal, Pomita and Roy, Samiran",
    title = "{Effect of Non Unitarity on Neutrino Mass Hierarchy determination at DUNE, NO$\nu$A and T2K}",
    eprint = "1609.07094",
    archivePrefix = "arXiv",
    primaryClass = "hep-ph",
    doi = "10.1016/j.nuclphysb.2017.04.018",
    journal = "Nucl. Phys. B",
    volume = "920",
    pages = "385--401",
    year = "2017"
}

@article{Escrihuela:2016ube,
    author = "Escrihuela, F. J. and Forero, D. V. and Miranda, O. G. and T\'ortola, M. and Valle, J. W. F.",
    title = "{Probing CP violation with non-unitary mixing in long-baseline neutrino oscillation experiments: DUNE as a case study}",
    eprint = "1612.07377",
    archivePrefix = "arXiv",
    primaryClass = "hep-ph",
    reportNumber = "IFIC-16-95",
    doi = "10.1088/1367-2630/aa79ec",
    journal = "New J. Phys.",
    volume = "19",
    number = "9",
    pages = "093005",
    year = "2017"
}

@article{Trzeciak:2025hap,
    author = "Trzeciak, Ana Maria Garcia and Nunokawa, Hiroshi and Quiroga, Alexander A.",
    title = "{Impact of unitarity violation on sensitivity of the leptonic CP phase at Hyper-Kamiokande and DUNE}",
    eprint = "2502.10873",
    archivePrefix = "arXiv",
    primaryClass = "hep-ph",
    doi = "10.1007/JHEP11(2025)059",
    journal = "JHEP",
    volume = "11",
    pages = "059",
    year = "2025"
}

@article{Antusch:2006vwa,
    author = "Antusch, S. and Biggio, C. and Fernandez-Martinez, E. and Gavela, M. B. and Lopez-Pavon, J.",
    title = "{Unitarity of the Leptonic Mixing Matrix}",
    eprint = "hep-ph/0607020",
    archivePrefix = "arXiv",
    reportNumber = "FTUAM-06-8, IFT-UAM-CSIC-06-30",
    doi = "10.1088/1126-6708/2006/10/084",
    journal = "JHEP",
    volume = "10",
    pages = "084",
    year = "2006"
}

@inproceedings{Hernandez-Garcia:2017pwx,
    author = "Hernandez-Garcia, Josu and Lopez-Pavon, Jacobo",
    title = "{Non-Unitarity vs sterile neutrinos at DUNE}",
    booktitle = "{Prospects in Neutrino Physics}",
    eprint = "1705.01840",
    archivePrefix = "arXiv",
    primaryClass = "hep-ph",
    reportNumber = "NUPHYS2016-HERNANDEZ-GARCIA, NUPHYS2016-LOPEZ-PAVON",
    month = "5",
    year = "2017"
}

@article{Huber:2007ji,
    author = "Huber, Patrick and Kopp, Joachim and Lindner, Manfred and Rolinec, Mark and Winter, Walter",
    title = "{New features in the simulation of neutrino oscillation experiments with GLoBES 3.0: General Long Baseline Experiment Simulator}",
    eprint = "hep-ph/0701187",
    archivePrefix = "arXiv",
    reportNumber = "TUM-HEP-656-07",
    doi = "10.1016/j.cpc.2007.05.004",
    journal = "Comput. Phys. Commun.",
    volume = "177",
    pages = "432--438",
    year = "2007"
}

@article{Huber:2004ka,
    author = "Huber, Patrick and Lindner, M. and Winter, W.",
    title = "{Simulation of long-baseline neutrino oscillation experiments with GLoBES (General Long Baseline Experiment Simulator)}",
    eprint = "hep-ph/0407333",
    archivePrefix = "arXiv",
    reportNumber = "TUM-HEP-553-04",
    doi = "10.1016/j.cpc.2005.01.003",
    journal = "Comput. Phys. Commun.",
    volume = "167",
    pages = "195",
    year = "2005"
}

@article{Hettmansperger:2011bt,
    author = "Hettmansperger, Hans and Lindner, Manfred and Rodejohann, Werner",
    title = "{Phenomenological Consequences of sub-leading Terms in See-Saw Formulas}",
    eprint = "1102.3432",
    archivePrefix = "arXiv",
    primaryClass = "hep-ph",
    doi = "10.1007/JHEP04(2011)123",
    journal = "JHEP",
    volume = "04",
    pages = "123",
    year = "2011"
}

@article{T2K:2025wet,
    author = "Abubakar, S. and others",
    collaboration = "T2K, NOvA",
    title = "{Joint neutrino oscillation analysis from the T2K and NOvA experiments}",
    eprint = "2510.19888",
    archivePrefix = "arXiv",
    primaryClass = "hep-ex",
    reportNumber = "FERMILAB-PUB-25-0132-PPD",
    doi = "10.1038/s41586-025-09599-3",
    journal = "Nature",
    volume = "646",
    number = "8086",
    pages = "818--824",
    year = "2025"
}

@article{Kaur:2021rau,
    author = "Kaur, Daljeet and Khan Chowdhury, Nafis Rezwan and Rahaman, Ushak",
    title = "{Effect of non-unitary mixing on the mass hierarchy and CP violation determination at the Protvino to ORCA experiment}",
    eprint = "2110.02917",
    archivePrefix = "arXiv",
    primaryClass = "hep-ph",
    doi = "10.1140/epjc/s10052-024-12431-3",
    journal = "Eur. Phys. J. C",
    volume = "84",
    number = "2",
    pages = "118",
    year = "2024"
}

@article{Singha:2021jkn,
    author = "Singha, Dinesh Kumar and Ghosh, Monojit and Majhi, Rudra and Mohanta, Rukmani",
    title = "{Optimal configuration of Protvino to ORCA experiment for hierarchy and non-standard interactions}",
    eprint = "2112.04876",
    archivePrefix = "arXiv",
    primaryClass = "hep-ph",
    doi = "10.1007/JHEP05(2022)117",
    journal = "JHEP",
    volume = "05",
    pages = "117",
    year = "2022"
}

@article{Raut:2009jj,
    author = "Raut, Sushant K. and Singh, Ravi Shanker and Sankar, S. Uma",
    title = "{Magical properties of 2540 Km baseline Superbeam Experiment}",
    eprint = "0908.3741",
    archivePrefix = "arXiv",
    primaryClass = "hep-ph",
    doi = "10.1016/j.physletb.2010.12.029",
    journal = "Phys. Lett. B",
    volume = "696",
    pages = "227--231",
    year = "2011"
}

@article{Dighe:2010js,
    author = "Dighe, Amol and Goswami, Srubabati and Ray, Shamayita",
    title = "{2540 km: Bimagic baseline for neutrino oscillation parameters}",
    eprint = "1009.1093",
    archivePrefix = "arXiv",
    primaryClass = "hep-ph",
    doi = "10.1103/PhysRevLett.105.261802",
    journal = "Phys. Rev. Lett.",
    volume = "105",
    pages = "261802",
    year = "2010"
}

@article{Denton:2025kvy,
    author = "Denton, Peter B. and Gehrlein, Julia and Kong, Chui-Fan",
    title = "{Testing new physics in oscillations at a neutrino factory}",
    eprint = "2502.14027",
    archivePrefix = "arXiv",
    primaryClass = "hep-ph",
    doi = "10.1016/j.nuclphysb.2025.117040",
    journal = "Nucl. Phys. B",
    volume = "1018",
    pages = "117040",
    year = "2025"
}

\end{document}